\documentclass[onecolumn]{aa} % for a paper on 1 column  
\usepackage{graphicx}
\usepackage{longtable}
\usepackage{supertabular}
\usepackage{txfonts}
\begin{document}

   \title{New BVRI photometry results on KBOs from the ESO VLT \thanks{Based on observations collected at the Very Large Telescope of the European Southern Observatory at Cerro Paranal in Chile under programs 072.C-0483(A) and (B)}}

   %\subtitle{}

   \author{P. Santos-Sanz
          \inst{1}
          \and
          J.L. Ortiz\inst{1}
		  \and
		  L. Barrera\inst{2}
		  \and
		  H. Boehnhardt\inst{3}}

   \offprints{P. Santos Sanz.}

   \institute{Instituto de Astrof\'{\i}sica de Andaluc\'{\i}a, 
   			 PO Box 3004, 18080 Granada, Spain.\\
	         \email{psantos@iaa.es}  
		 \and
            Universidad Metropolitana de Ciencias de la Educaci\'on,
			Facultad de Ciencias B\'asicas, Dpto. de F\'{\i}sica, Santiago de Chile, Chile.\\
		\and
			Max-Planck-Institute f\"ur Solar System Research,
			 Max-Planck-Str. 2, D-37191 Katlenburg-Lindau, Germany.
            }

   %\date{Received ; accepted , }

% \abstract{}{}{}{}{} 
% 5 {} token are mandatory
 
  \abstract
{Photometric surveys of transNeptunian objects (TNOs) and Centaurs have
suggested possible correlations between some orbital parameters and surface
colors of classical objects, scattered disk objects (SDOs), and Centaurs.
However, larger sample sizes are needed in order to corroborate or rule out
the possible correlations and find some possible new ones. The implications
from these correlations for the formation and evolution of classical TNOs,
SDOs, and Centaurs are important to understand possible physico-chemical
coloring mechanisms and their influence on the surfaces of the TNOs and
Centaurs, as well as their evolutionary history.}
%_______________________________________________________________________________
{We aim to obtain a sufficiently large photometric dataset
in order to carry out a more significant statistical analysis, with emphasis on
improving the sample size of observed objects, in particular for classical
TNOs and SDOs.}
%_______________________________________________________________________________
{We use VLT-FORS images through BVRI filters of 32 Kuiper Belt Objects (KBOs)
and obtain their colors after proper reduction and calibration. We study
the possible correlations merging these new measurements with the VLT
published results from the ESO large program and with the latest published results of the Meudon Multicolor
Survey via non-parametric
statistical tests.}
%_______________________________________________________________________________
{We obtain a large dataset of 116 objects (classical, SDOs and Centaurs) and, in addition to confirming most of the correlations and conclusions reached in the literature, some possible
new correlations are found. The most interesting ones are some correlations
of color vs. orbital parameters for the different dynamical groups. We find
that some correlations in the classical group, as well as the (dynamically) cold and hot subgroups depend on the size of the objects. As a by-product of our
study, we were able to identify new candidates for light curve studies and found 
that $\sim$ 55\% of the objects showed variability above 0.15 mags. This is a higher value than what is found in other studies. Since our sample contains smaller objects than samples from other studies, this result might be an indication that the smaller TNOs are more elongated than the larger ones.}
%_______________________________________________________________________________
{} 
  

   \maketitle
%
%_______________________________________________________________________________

\section{Introduction}\label{intro}

Trans--Neptunian objects (TNOs) are cold minor bodies composed of ice and
rock that populate the region of the solar system beyond Neptune. Their existence
was hypothesized to explain the observed distribution of short-period comets (\cite{fern80}; \cite{dunc88}). The first of these 
objects, besides Pluto, was discovered in 1992 by Jewitt \& Luu
(1993), and currently more than 1000 of these objects have been 
detected. 

%_______________________________________________________________________________

\textbf{Dynamical groups:} The known TNOs comprise different dynamical groups:
classical objects which exhibit orbits with moderate eccentricities; resonant objects trapped by Neptune in mean motion resonances (the \emph{Plutinos} being the most
representative population, in the 3:2 resonance); scattered disk objects
(SDOs) with orbits with high eccentricities and sometimes high 
inclinations, due to close encounters with the planet Neptune in the 
past. The classical group is sub-classified in two sub-groups: The 
so-called (dynamically) \emph{hot} classical objects with orbital 
inclinations $>4.5\,^{\circ}$, and the (dynamically) \emph{cold} 
classical objects, with inclinations $<4.5\,^{\circ}$. Centaurs 
represent TNOs scattered towards the inner solar system and they 
reside between the orbits of Jupiter and Neptune, so many investigators 
consider them Kuiper Belt Objects (KBOs). The current scenarios of the Solar 
System formation suggest that the KBOs, the short-period comets and 
some icy satellites of the major planets were formed in the 
transNeptunian belt. What is known of their physical nature supports 
this idea. However, since the TNOs never appear in the vicinity of the 
Sun, they are believed to be even more pristine than their relatives. 
The knowledge of the physico-chemical nature of TNOs may give important 
clues on the conditions prevailing at the time of their formation, 
mainly for the larger TNOs (some of which comply with the dwarf-planet 
definition) which are the least collisionally evolved objects.

%_______________________________________________________________________________

\textbf{Colour census of TNOs:} Over the past years the sample of TNOs 
with BVRI colors measured has grown to more than 100 objects. Statistical
analyses suggest the presence of various correlations among the color and
dynamical properties of classical TNOs and SDOs, while Plutinos and 
Centaurs appear to be heterogeneous and without obvious relations 
between physical and dynamical parameters of the objects. From BVRI 
color studies Tegler \& Romanishin (2000) suggested the existence of 
a group of red classical objects beyond 41 AU. Based upon larger 
data sets other authors (\cite{Boe02}; \cite{Doress}) provide much 
clearer evidence for this first TNO group that could be identified 
from similar physical and dynamical properties: classical objects 
of very red color (spectral gradients above $\sim$30\%/100 nm) in 
dynamically cold orbits (low inclinations to the ecliptic and low 
eccentricities).

Trujillo \& Brown (2002) find a correlation between inclination and 
surface reddening among classical TNOs and SDOs: the objects tend 
to be bluer the higher their inclinations. This trend is also shown 
in the work of Doressoundiram et al. (2002) with the additional 
emphasis that the high inclination objects are intrinsically brighter 
(possibly larger) than the dynamically cold ones. Both 
trends are based on a statistical method that gives higher weight 
to objects with properties close to the trend line. Consequently, 
the claimed correlations appear much weaker in normal linear regression 
fits as performed by Boehnhardt et al. (2002). In addition, they find in the same data set a stronger correlation between the 
color range of classical objects and perihelion distance: More 
explicitly, a linear decrease of the spectral gradient with decreasing 
perihelion distance may exist for classical objects with perihelion 
between 40 and 36 AU. The correlation coefficients of that trend 
appear to be higher than those involving inclination. However, at 
present they are not significant enough for either of the abovementioned correlations to be considered ``certain''. Moreover, some obvious 
outliers exist for each of these trends. On the other hand, it appears 
that the overall spectral gradient range of TNOs (for classicals, SDOs, 
Plutinos) and Centaurs is confined to -10 and 55\%/100 nm and is fully 
populated for objects that come closer to the Sun than $\sim$35 AU.

%_______________________________________________________________________________

\textbf{Surface evolution scenarios:} The standard scenario for the surface 
evolution of TNOs invokes the counteraction of mainly two effects: (1) 
Chemical alteration due to high-energy radiation and (2) resurfacing by 
icy deposits after collisions. The first process is expected to produce 
a surface crust of reddish color, while the latter should cause bluish colors. 
Some modelling results (\cite{Thebault03a}), though using simplified 
assumptions on the transNeptunian belt structure and the physical 
resurfacing processes, do not support the abovementioned color 
trend with inclination. Instead, it predicts a correlation with 
eccentricity that is not found in the existing data. Gomes (2003) and 
Morbidelli et al. (2003) introduced an interpretation for the excited 
classical population (classical objects in orbits of higher inclination 
and eccentricity, the hot population): These objects could be scattered 
bodies from the region of the gas giants Neptune and Uranus. Due to the 
different environment closer to the Sun, these objects may have grown 
larger before they were injected in the classical region, where they 
account for the otherwise unexplained hot population of objects. The authors suggest, as 
a by-product that the chemical properties of the 
scattered classical objects could be different from those of the cold 
classicals, which should explain the trend of bluer instead of very red 
spectral gradients. However, apart from dynamical evidence for the 
existence of a hot classical population, the arguments for their 
physical and chemical diversities appear to be of an ad-hoc nature. 
The color trend with perihelion distance together with further evidence 
(atmosphere around Pluto, lightcurve change in 1996TO$_{66}$) has 
led other authors (\cite{Hainaut00}; \cite{Boe01}, and 2002) to propose 
a third resurfacing effect for TNOs: icy recondensation of gas and deposit 
of haze from a coma-like atmosphere that is produced by intermittent 
intrinsic activity of the object. The fresh surface ice would appear 
bluer. The perihelion, aphelion or semimajor axis dependence could be 
seen as an indicator of the increasing activity level of the agent 
driving the activity when the KBOs move closer to the Sun: N$_{2}$ and 
CO ice sublimation could be a possibility (\cite{Delse82}). Nevertheless, 
the lightcurve change in 1996TO$_{66}$ may also be explained as being due to
an opposition effect (\cite{Belska06}).

%_______________________________________________________________________________

\vspace{0.43cm}
The scientific aim of our new observations is that, in order to increase the
TNO/Centaur 
sample size, we present in this paper colors for 32 KBOs. In particular, 
the SDO and classical sample increase significantly. 
We merge these measurements with other published results in order to 
allow a large and unique dataset of objects that we analyse with different 
statistical methods in order to obtain relevant physical information.

%_______________________________________________________________________________

\section{Observations and data reduction}

The observations presented here are a continuation of the
Large Program developed for the study of the physical properties of TNOs and Centaurs executed at the European Southern Observatory between 2000 and 2002
(Boehnhardt et al. 2002). Our new observations were performed during 4
nights; 1st and 2nd November 2003 and 25th and 26th March 2004, with the Very
Large Telescope (VLT) unit telescope Antu (8.2 m) at Cerro Paranal, Chile (ESO)
using FORS1 focal reducer instrument (detector Tek CCD 2048 x 2048 pixels;
FOV= 6.8 x 6.8 $\rm{arcsec}$; resolution = 0.2 $\rm{arcsec}$/pixel). The
imaging through Bessell broadband BVRI filters consisted of RVIR exposure 
sequences of $\sim$1h maximum duration to correct for brightness variations 
due to rotation. Preference was given to fainter KBOs (23-24mag) in order 
to compensate the bias towards brighter (most likely larger) objects in 
the current color sample. For brighter targets we also took B filter
photometry (in filter sequences RBRVIR). In most of the cases the observed 
objects have well established orbits (2-6 oppositions are measured) except 
for 2003QB$_{112}$ that had - by then - an observed orbit arc of only 57 days
(Table \ref{tableOBS}).  

The images were taken under photometric conditions and dark skies. The
telescope was tracked at sidereal rate with exposure times between 180-1200 
seconds depending on the object brightness (no differential tracking 
was needed due to the slow velocity of the KBOs $\sim$1 arcsec/h). We choose 
long enough exposure times to obtain a reasonable signal-to-noise-ratio for 
the objects ($>$25). The phase angles of the objects were in the 
range 0.1-1.2 deg in order to observe with the optimal object reflectivity 
position (near opposition) and to avoid possible reddening effects due to 
larger phase angles (\cite{Millis76}; \cite{Bowell79}).

%_______________________________________________________________________________

\textbf{Data reduction and photometry:} Each image was reduced using standard
techniques of bias subtraction and flat field correction. Landolt standard stars at different airmasses for each night were used in order to calculate the photometric calibration 
parameters (photometric zero points, first order extinction coefficients
and color term corrections). 

%No cosmic ray removal algorithms  were used and we rejected the images in which a  cosmic ray hit or a star was close to the  object.

%_______________________________________________________________________________

As the selected TNOs and Centaurs are faint (m$_{R}\sim$20-24 mag.) we
measured their magnitudes using small apertures and correcting for the 
flux loss by means of the aperture correction technique. To carry out 
this correction, we use 30 unsaturated stars in each image field in order 
to compute the aperture correction for each photometric object measurement 
with high accuracy. The object aperture radius varied between 2-6 pixels 
depending on the night conditions and the brightness of the object. 
We chose the aperture radius that gave the maximum signal-to-noise-ratio for 
each object by calculating the curve of growth for a number of stars per 
image (Howell, 1989). An aperture area of $\sim$25 pixels was found to give
the smallest changing correction from image to image. We compute the 
total error in the photometric process as:

\begin{equation}
\sigma =\sqrt{\sigma_\mathrm{phot}^{2}+\sigma_\mathrm{apcorr}^{2}+\sigma_\mathrm{calib}^{2}}
\end{equation}

where 

$\sigma_\mathrm{phot}$ is the error of the photometric measurements 
as estimated from Poisson statistics and using the appropriate CCD
parameters.

$\sigma_\mathrm{apcorr}$ is the uncertainty of aperture correction 
resulting from the dispersion among measurements of the field stars. This term includes intrinsically systematic errors like flatfielding gradients and variations.

$\sigma_\mathrm{calib}$ is the calibration error. This error is the quadratic sum of the rms of the standard stars fit, and of the derived errors of the zeropoints, extinction coefficients and color-term determinations.

%_______________________________________________________________________________

\subsection{Photometry of the new sample}

\textbf{Overview and comparison with data from other authors:} The main color 
and photometric results of the new sample ordered by the object's designation 
are listed in Table \ref{tableRES}. For the objects that show short-term brightness variability in the R-filter (see Table \ref{BrightVAR}) we compute the mean value of all R-magnitudes of the sequence in order to obtain the color indexes. 13 of the 32 objects had colors measured by other authors: 1997SZ$_{10}$ (\cite{Tegler00}); 2000CM$_{114}$, 
2000CR$_{105}$, 2001FM$_{194}$, 2001QX$_{322}$, 2001SQ$_{73}$
(\cite{Tegler03}); 2000CN$_{105}$, 2000YW$_{134}$ (\cite{Peix});
2001UR$_{163}$, 2002GH$_{32}$, 2002GJ$_{32}$ (\cite{Doress_b}); 2002UX$_{25}$ (\cite{Rabin07}); and 2003AZ$_{84}$ (\cite{Fornasier04}). 5 of these 13 objects show differences in 
colors and/or magnitudes with respect to our results: 2000CR$_{105}$, 
2001FM$_{194}$, 2001QX$_{322}$, 2001UR$_{163}$, and 2002GJ$_{32}$. 
The differences in colors of 2000CR$_{105}$ and 2001QX$_{322}$ are $< 2
\sigma$. For 2000CR$_{105}$ it may be explained by the brightness 
variability (Table \ref{BrightVAR}). For the other 3 objects the 
differences in magnitudes or colors are $> 2 \sigma$ and may arise 
from different observational circumstances (different dates, phase 
angles, etc.), calibration errors, or possible brightness variability 
due to rotation.

The remaining 19 object measurements are new. Henceforth, `OWN' indicates 
our own photometric results, `LP' are the VLT published results of the 
ESO large program (\cite{Boe02}; \cite{Peix}), and `2MS' are the Meudon Multicolor Survey photometric 
results (\cite{Doress_b}). We use our own color results in the computations of OWN+LP and 
OWN+LP+2MS data set in the cases when our objects were repeated 
measurements in LP , or 2MS.

%_______________________________________________________________________________

%_______________________________________________________________________________

\begin{figure}
   \centering
   
   \includegraphics[angle=-90,width=9.07cm]{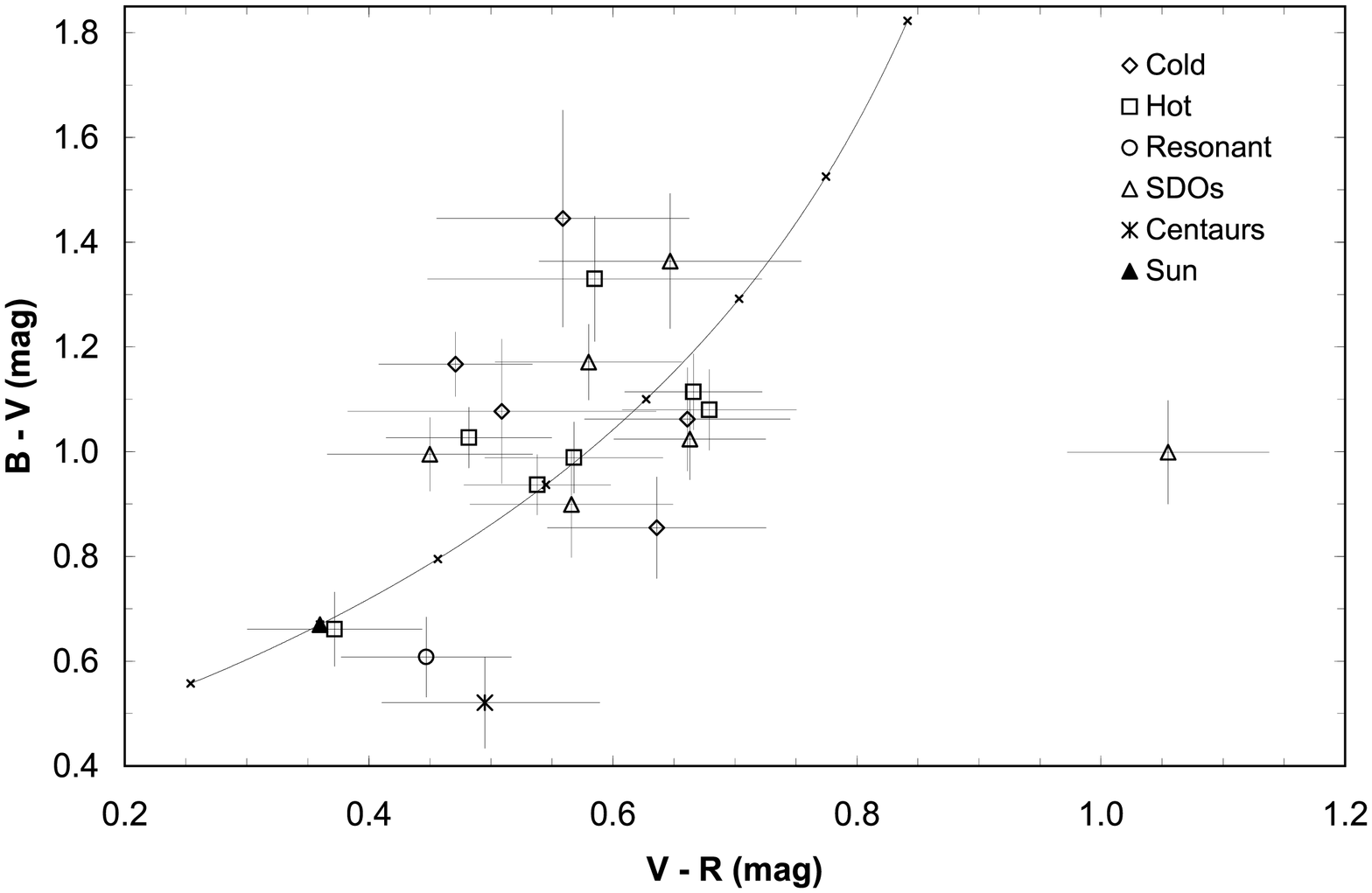} 
   \includegraphics[angle=-90,width=9.0cm]{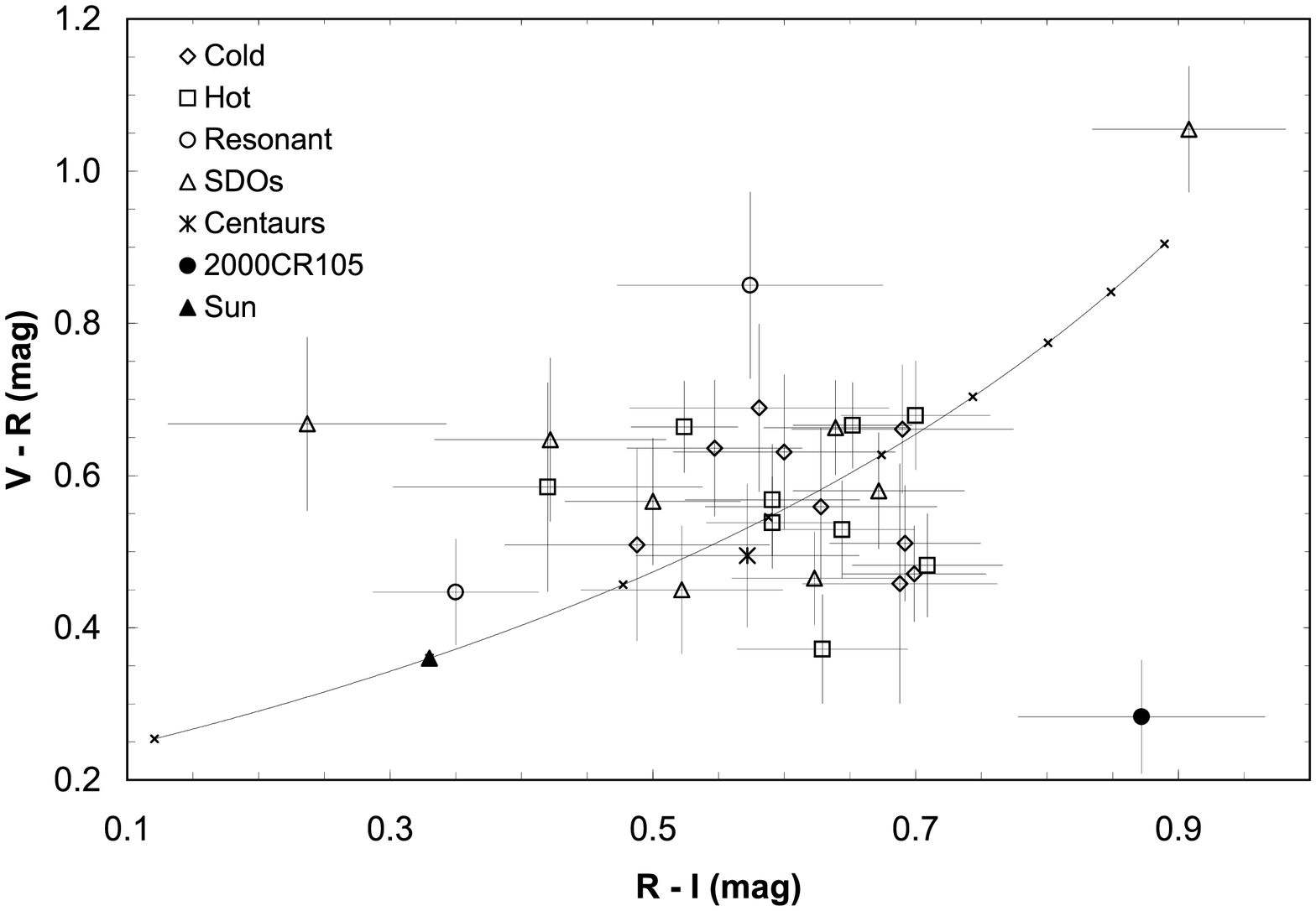}
   
      \caption{V-R vs. B-V (\emph{top}) and R-I vs. V-R (\emph{bottom}) for
the whole new sample. Symbols: classical cold objects (\emph{open diamonds}),
classical hot objects (\emph{open squares}), Neptune{'}s resonant objects (open circles), SDOs (\emph{open triangles}), Centaur (\emph{asterisk}), 2000$CR_{105}$ (\emph{filled circle}) and the Sun (\emph{filled triangle}). The reddening line, which is the locus of objects
displaying a linear reflectivity spectrum, has a range of spectral gradients
from -10 to 70\%/100 nm (in the top figure only from -10 to 60\%/100 nm);
a cross mark is placed at every 10 units.}

         \label{colors_own}

   \end{figure}

\textbf{Color-color diagrams:} The color-color plots for our sample 
are shown in Fig. \ref{colors_own}. The top panel shows V-R versus B-V. The error bars and different symbols for classical cold objects (\emph{open
diamonds}), classical hot objects (\emph{open squares}), Neptune{'}s resonant objects (\emph{open circles}), SDOs (\emph{open triangles}), Centaur (\emph{asterisk}) and the Sun (\emph{filled triangle}). 

The reddening line is drawn. This line is the locus of objects 
displaying a linear reflectivity spectrum and has a range of spectral 
gradients from -10 to 60\%/100 nm (very red). 

There are approximately the same number of points above and below the reddening line. This indicates a roughly constant spectral slope in the BVR range. One outlier is apparent: 2001UR$_{163}$ (\emph{right open triangle}), which is an SDO. This object was 
measured by Doressoundiram et al. (2005) who obtained redder 
values than ours. We obtain, nevertheless, that 2001UR$_{163}$ is 
a very red object  (B-R=$2.05\pm 0.10$), not far from the result 
by Doressoundiram et al. (B-R=$2.28\pm 0.04$). The difference may 
be due to rotational color variations, and never due to the opposition 
effect (because the observing phase angles are the same ($\alpha=0.3$) 
for both measurements). Nevertheless, Sheppard \& Jewitt (2003) find no measurable 
photometric variations, with lightcurve amplitude $<0.08$ magnitudes 
and/or period $>$24 hours. Hence, the different results for this particular
object remain unexplained, except if we assume an underestimation of the given errors, perhaps of the calibration ones.

The bottom panel of Fig. \ref{colors_own} shows R-I versus V-R, using the 
same symbols as for the objects in the top panel, now adding the distant 
TNO 2000$CR_{105}$ (\emph{filled circle}) which is considered to be a member 
of the extended scattered disk. The reddening line is drawn from -10 to 
70\%/100 nm. The number of points lying above the reddening line is slightly greater than points lying below, which is an indication of a slight increase of the spectral slope over the VRI range. Two outliers are apparent: 2001UR$_{163}$ (\emph{top right open 
triangle}) and 2000$CR_{105}$ (\emph{bottom right filled circle}). For the
color difference of 2001UR$_{163}$ see the previous paragraph. The color differences 
of 2000$CR_{105}$ to the overall trend may be explained 
because this object is a member of the poorly known extended scattered 
disk population (\cite{Gladman02}) which contains objects with orbits 
decoupled from Neptune. However, no real evidence for this
speculative statement can be provided.

%_______________________________________________________________________________   
   
\subsection{Objects showing short-term brightness variability}

From the total sample of 32 objects we have 11 objects with some temporal 
coverage in the R filter to study a possible short-term rotational variability. 
For the analysis described below we take into account objects that show R
magnitude variations $\geq 0.15$mag during the exposure sequence (this value
is larger than the mean measurement error listed in Table \ref{tableRES}). 
Among these 11 objects, 7 present short-term variability greater than three 
times the relative photometric error bars (3$\sigma$ variations), and 4 
present possible short term variability equal to or below two times the error 
bars ($\leq 2\sigma$) as is shown in Table \ref{BrightVAR}. Here, we use as
the error reference $\sigma = \sigma_\mathrm{phot}$, i.e. we do not take into 
account aperture correction ($\sigma_\mathrm{apcorr}$) or calibration errors ($\sigma_\mathrm{calib}$), 
because we observe each object during only one night, and we are computing magnitude variations ($\Delta$Rs) for small temporal intervals ($\Delta$ts) therefore, under these particular conditions, $\sigma_\mathrm{apcorr}$ and $\sigma_\mathrm{calib}$ are negligible in $\Delta$Rs.

2 of 11 objects ($\sim18$\%) show variations $\geq$0.40 mags., a rate similar
to that obtained by Ortiz et al. 2006 ($\sim$16\%) and Sheppard \& Jewitt
2003 ($\sim$15\%). 4 objects ($\sim$36\%) show variations between 0.15 and 0.40
mag, a percentage higher than that by Ortiz et al. 2006 ($\sim$15\%) and 
Sheppard \& Jewitt 2003 ($\sim$12\%). In total 6 objects ($\sim$55\%) 
display variations $\geq$0.15 mags., again more than found by Ortiz et al. 
2006 ($\sim$31\%), Sheppard \& Jewitt 2003 ($\sim$27\%) and \cite{Lacerda06} 
($\sim$30\%). We must take into account too that the average temporal
coverage for the eleven objects is $\sim$1 hour, a value lower than the 
typical rotation period for the TNOs (likely to be of the order of a few to
several tens of hours). Hence, the amplitude values listed in Table
\ref{BrightVAR} are likely lower limits, which implies that more than 55\% of
the objects may have amplitude variations $\geq$0.15 mag. This value is
higher than those by other authors. However, considering that our objects are
significantly fainter than those studied elsewhere, this trend may not be a surprise, because
smaller sized objects are expected to be more distorted by collisions than the larger
ones, as already pointed out by Lacerda \& Luu (2006). Although less likely there is also the possibility that the photometry errors have been underestimated.

For the colour estimation of the objects that show R-brightness variations,
we calculate a time interpolated R-value for the epoch of the respective complementary filter
measurement in the exposure sequence. This interpolated value is the R-magnitude shown in Table \ref{tableRES}.

%_______________________________________________________________________________

\subsection{Absolute magnitudes}

We compute the absolute magnitudes for all the objects using two different 
methods: the Bowell formalism (\cite{Bowell89}), and the linear phase function
approximation. For the Bowell formalism, we use the so-called G parameter. The 
G value is obtained as an average value weighted with the errors of the G values for TNOs published by Sheppard \& Jewitt (2003). In doing so, we obtain $G= -0.03\pm 0.02$ for the TNOs. Moreover, we correct the measured magnitude (mag.) for the distances object-sun ($r$) and object-earth ($\Delta$), to obtain $H_{\alpha}$,

\begin{equation}
	H_{\alpha}=mag.-5\cdot log(r\cdot\Delta)
\end{equation}

Using the average G value, the phase angle $\alpha$, and the Bowell 
parameters: A$_{1}$=3.33, A$_{2}$=1.87, B$_{1}$=0.63, B$_{2}$=1.22, we apply
the phase correction of the Bowell formalism (\cite{Bowell89});

\begin{equation} 
	\varphi{_1} = e^{(-A_{1}\cdot tan(\alpha)^{B_{1}})}
\end{equation}

\begin{equation}
	\varphi{_2} = e^{(-A_{2}\cdot tan(\alpha)^{B_{2}})} .
\end{equation}
	
Then, the absolute magnitude, H, is derived as (\cite{Bowell89}):

\begin{equation}
	H=H_{\alpha}+2.5\cdot log ((1-G)\cdot\varphi_{1}+G\cdot\varphi_{2}) .
\end{equation}

On the other hand, we also compute the absolute magnitude using the linear
phase function approximation as:

\begin{equation}
	H=mag.-5\cdot log(r\cdot\Delta)-\alpha\cdot\beta
\end{equation}

where mag., r, $\Delta$, and $\alpha$ are the same parameters as explained 
above and $\beta$ is the phase curve slope or \emph{linear} phase
coefficient. For TNOs we take $\beta=0.16\pm 0.03$ mag/deg, and for 
Centaurs $\beta=0.11\pm 0.01$ mag/deg (modal values referring to the 
$\beta$ estimations by \cite{Sheppard02}).

For each object we compute the V and R absolute magnitude using 
the two formalisms above. The four resulting absolute magnitudes 
($H_{V}(Bowell)$, $H_{V}(Linear)$, $H_{R}(Bowell)$, and $H_{R}(Linear)$) are
listed in Table \ref{AbsMag}. The absolute magnitudes used for the
statistical calculations are given under column $H_{R}(Linear)$, ensuring in this way
the compatibility of our data with those by Doressoundiram et al. (2002) and 
Peixinho et al. (2004).

%_______________________________________________________________________________

\subsection{Spectral gradients}
\label{Grt}

It is known that the TNO and Centaur spectra are approximately featureless
in the visible and display almost constant slopes over a wide wavelength
range. This is also reflected in the color vs. color plots of our sample 
(Fig. \ref{colors_own}). However, the slope of TNO 
spectra is not really constant over the full visible spectrum (at least 
in a certain number of cases); most noteworthy is the slope change towards 
the near-IR (\cite{Davies00}; \cite{Del06}). Hence, using a set of 
filters over the full wavelength range for slope fitting can provide quite 
accurate results, with the possible disadvantage that slopes of some objects
are affected by the change-over in the very red, but some others are not affected. 
In our case we compute the spectral gradients -Grt- (the slopes) for 
the colors B-V, V-R and R-I. Taking the average of Grt$_{(B-V)}$, 
Grt$_{(V-R)}$, and Grt$_{(R-I)}$, we obtain a single Grt value per object 
that includes all the color information in the [436,797 nm] wavelength 
range (Table \ref{tableRES}). Hence, first we obtain the partial gradients, Grt$_{(\lambda_{1}-\lambda_{2})}$, using the formula (\cite{Hainaut}):

\begin{equation}
\label{Grt_equation}
	Grt_{(\lambda_{1}-\lambda_{2})}=\frac{R(\lambda_{1})-R(\lambda_{2})}{|\lambda_{1}-\lambda_{2}|}\cdot 10^4
\end{equation}

where $\lambda_{1}, \lambda_{2}$ are the central wavelengths of the different filters (B,V,R, and I) expressed in nanometers (nm) and $R(\lambda_{n})$ are the relative spectral reflectivities (\cite{JewittMeech86}) normalized to 1 in the V-filter central wavelength and expresed as:    

\begin{equation}
\label{Reflectivity}
	R(\lambda_{n})=10^{-0.4\cdot [(m(\lambda_{n})-m(v))-(m(\lambda_{n})_{\odot}-m(v)_{\odot})]}
\end{equation}

where $m(\lambda_{n}), m(v)$ are the magnitudes of the object in $\lambda_{n}$ and V
filters, $m(\lambda_{n})_{\odot}, m(v)_{\odot}$ are the magnitudes of the sun in the same filters. Thereafter, error weighted averaging is performed to obtain the mean value Grt.

The results derived for Grt are given in \emph{percent per 100 nm}  (\%/100
nm). If measurements in all the filters (B,V,R,I) exist, we average the
mean value using three partial Grt of adjacent filter colors; if only three
filters data are available, the mean value is derived from two partial Grts, 
etc. The gradient results for our object sample are shown in Table
\ref{tableRES}. We consider them to be a good indicator of the gradient 
slopes over the B-I wavelength range. The gradient errors in the Table 
are computed using the statistical errors of the average colors, and 
the errors due to the spectral gradient average. 

\begin{figure}
   \centering
   \includegraphics[angle=-90,width=8.8cm]{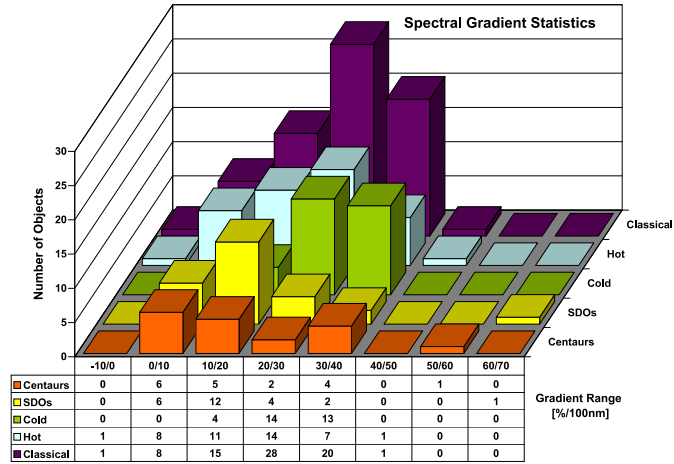}
   \caption{Reddening gradient statistics of TNOs and Centaurs for the
OWN+LP+2MS sample. Dynamical classes: classical, hot, cold, scattered disk
objects (SDOs), and Centaurs. The reddening of the objects (in \%/100 nm) as
compared to the sun is calculated using Eq. \ref{Grt_equation} and the number
of objects (Y axis) per reddening interval (X axis) is counted.}  
        \label{Histogram}
   \end{figure}

We compute the mean values of spectral gradient and the slope distribution
for each dynamical group of TNOs in our own sample using the full datasets:
OWN+LP+2MS. We use these results to seek possible correlations of 
spectral gradients with orbital parameters and H$_{R}$ (see Table
\ref{Grt_correl} below). These results will be discussed in 
Section \ref{DataAnalysis} and subsections thereof. As can be seen in
Fig. \ref{Histogram} and in Table \ref{Avg_Grt} there are characteristic
differences between the red cold classical cluster, with a high average 
reddening gradient (Grt=27.4\%/100 nm), and the hot classical objects, 
with a low average reddening gradient (Grt=19.6\%/100 nm), but we can also note that both reddening gradients are compatible within the error bars. Despite possibly
different slope distributions, the mean spectral gradients of SDOs and
Centaurs agree within errors with that of the hot classicals. Our results 
are compatible with those obtained by Hainaut \& Delsanti (2002); Boehnhardt et al. (2002); and Boehnhardt et al. (2003).

%______________________________________________________________________________
   
\section{Statistical analysis and discussion}
\label{DataAnalysis}

In order to search for correlations between different colors and orbital 
parameters (a, semimajor axis; q, perihelion distance; Q, aphelion 
distance; e, eccentricity; i, inclination) we use non-parametric statistical 
tests because these methods do not assume any particular population 
probability distribution, nor any functional shape. In particular we 
compute all the correlations using the Spearman rank correlation, $\rho$ 
(Spearman, 1904). This method is distribution-free and less sensitive to 
outliers than other ones. We studied the strength of the correlations by 
computing the Spearman coefficient $\rho$ and the significance level (SL) 
as the probability that the null hypothesis (samples not correlated) is 
not true. We consider that correlations are strong when $\rho$ is greater 
than 0.6, a possible weak correlation exists when $\rho$ values are between 
0.3 and 0.6, and no correlations are present for values less than 0.3. 
Following Efron and Tibshirani (1993) we use the qualification
criteria: SL greater than 95\% = reasonably strong evidence of correlation; 
SL greater than 97.5\% = strong evidence of correlation; SL greater than 
99\% = very strong evidence of correlation. The interpretation of $\rho$, 
SL, and the graphical representation of the magnitudes allows us to detect 
possible correlations among the different magnitudes. Apart from these qualification
criteria we consider that a possible correlation could exist when $\rho$ is greater or equal than 0.3 and SL is greater than 80\%, in particular for samples with few objects. We emphasize that this particular threshold ($\rho \ge$ 0.3; SL$>$ 80\%) is only a hint of a possible correlation that would need more data to be confirmed. First, we correlate only the sample of our new results (OWN), next we merge our objects with the results of the ESO Large Program 
(LP; \cite{Boe02}; \cite{Peix}) and, finally, we add the results of the Meudon Multicolor 
Survey (2MS; \cite{Doress_b}). A summary of the number of objects for each 
group and sample can be seen in Table \ref{Summary}.

The Spearman statistical analysis of the results is made by separating 
the dataset in different dynamical subgroups, ie: classical objects, 
hot and cold classical objects, scattered disk objects (SDOs), Centaurs 
(Cent) etc. We use the classification scheme described in Doressoundiram et al. (2005). Centaurs and SDOs are merged on the MPC list (http://www.cfa.harvard.edu/iau/lists/Centaurs.html) with no distinction. From this list we consider Centaurs the objects with perihelia (q) below 30 AU (Neptune{'}s semi-major axis) and aphelia (Q) below 48 AU (2:1 resonance). The other objects are classified as SDOs\footnote{The only exception is the very distant object 2000CR$_{105}$ with Q= 394 AU and a= 219 AU, which is classified as an extended scattered disk object (ESDO).}. Using the MPC TNOs-list (http://www.cfa.harvard.edu/iau/lists/TNOs.html) we consider as classical objects all non-resonant objects between the 3:2 and 2:1 resonances ($39.5 AU <a<48 AU$) and outside Neptune{'}s Hill sphere (q$>35 AU$). Objects with low inclined orbits, q$< 36 AU$ and 36 AU$<$a$<$40 AU are also stable and classified as classical objects (Duncan et al. 1995). Two objects are in mean motion resonance with Neptune: 1997SZ$_{10}$ (2:1 resonance) and 2003AZ$_{84}$ (3:2 resonance, Plutino). We do not include these two objects in our statistical analysis. We merge some subsets that could be physically or evolutionally related 
in order to search for a possible increase in the correlations of the 
merged sets, ie: SDOs+hot, SDOs+Cent. In some cases, color information 
is not available for some objects (ie: in B-filter), for this reason we 
always indicate the number of objects in each correlation computation. 
The corresponding correlation coefficients and significance levels are 
listed in Table \ref{correl}. Here, we discuss only the most relevant 
results. 

%_______________________________________________________________________________

\subsection{Color-color correlations}

\label{color-color} 

Correlations of colors are to be expected if one takes for granted the conclusions from
measurements of visible spectra of TNOs and Centaurs (see ESO LP papers: \cite{Boe02}; \cite{Boe03}; \cite{Del06}; \cite{Peix}), i.e. the spectra are generally
featureless and have a rather constant slope with only moderate deviations 
to smaller gradients in the red. Hence, color-color correlations can test the
ensemble properties in this respect, i.e. whether or not the general picture
of the spectral reddening of the TNOs and Centaurs is consistent for the
objects of the sample. In particular, colors covering a wide wavelength range should fulfill the good correlation criteria. Correlations of ``neighbouring'' colors may be less prominent, since measurement errors can play a greater role. Anti-correlations of colors would be of real interest since they could indicate a unique absorption (less likely emission) feature commonly present in the visible spectra of the ensemble of objects.  

The inspection of the correlation parameters for the various object samples,
partially included in Table \ref{correl}, indicates a general
agreement with expectations, i.e. good color-color correlations exist for the
dynamical classes of classicals (hot and cold ones), SDOs, and Centaurs. 

In the following discussion of correlation properties of TNOs and Centaurs, we will not mention explicitly any color-color correlations.

We extended this analysis to the correlation of colors versus the 
$\psi$ parameter, which is a measure of the average energy of the 
collisions undergone by a TNO or Centaur. $\psi$ is proportional 
to the square of the average relative velocity of the object with 
respect to a circular orbit and depends on the orbital inclination (i), the eccentricity (e), and the semimajor axis (a) 
through (\cite{Opik76}):

\begin{equation}
	\psi=\frac{sin^{2}i+0.625\cdot e^{2}}{a}  .
\end{equation}

%____________________________________________________________________________   

\subsection{All classical objects (n=73)}
\label{classical} 

For the objects of the classical transNeptunian belt we confirm almost 
all the correlation results and trends found in previous
publications (\cite{Peix}; \cite{Doress_b}). However, we are able to identify 
some new correlations and differences (Table \ref{correl}). Our own sample of
classicals listed in Table \ref{tableOBS} amounts to about 26\% of the 
total number of classicals with colors measured (OWN+LP+2MS). 

\textbf{Strong correlations:} The very strong correlations are 
obtained for colors B-R and B-V vs. inclination i (see Table \ref{correl}) as well as spectral gradient Grt vs. i (see Table \ref{Grt_correl}). This finding is in agreement with results by other authors (\cite{Peix}; \cite{Doress_b}). We also obtain very strong evidence (SL$>$99\%) for the B-R vs. q and B-V vs. q correlations. Correlations between colors or spectral slopes and the dynamical orbit parameters i, q and e are usually discussed to validate or rule out the collisional resurfacing enviroment (Thebault and Doressoundiram, 2003). Despite the fact
that such correlations exist, their interpretation for the surface evolution of the classical TNOs remains unclear since the model predictions are indeed controversial (see section \ref{intro} and references therein). 

%_______________________________________________________________________________

%_______________________________________________________________________________

\textbf{Weaker correlations:}\label{weaker_classicals} Here we analyze correlations with low $\rho$s, but with strong SL values. These correlations are useful for the study because they imply broad but significant trends. Apart from this, we use the threshold $\rho \ge$ 0.3 and SL$>$ 80\% as a hint to identify possible correlations, which might be more studied in the future (see section \ref{DataAnalysis}). We find three interesting correlations (Tables \ref{correl} and \ref{correlsizes1}); R-I vs. q, that is not obvious from previous analyses, and two were already known, i.e B-R vs. e, and V-R vs. q. 

R-I vs. q displays evidence of a very strong correlation (n=69, $\rho$=0.30, 
SL=98.71\%) which supports the scenario that the redder objects have orbits 
with larger perihelion distance. This is consistent with the already known 
strong B-R vs. q correlation.

Regarding B-R vs. e, i, and q we can say that the classicals present (1)
higher inclinations the bluer they are, (2) shorter perihelion distances and higher 
eccentricity. These objects are usually referred to as the dynamically hot
population. Vice versa, the red classicals have lower inclinations, 
greater perihelion distances and lower eccentricity orbits; they represent the
dynamically cold population. The short perihelion distances of the bluer 
objects could support a coloring scenario by resurfacing with fresh 
``bluish'' ices due to sublimation of rare volatile species (Delsemme, 1982) 
around perihelion that freeze out subsequently when the object is further 
from the Sun. Other coloring mechanisms may explain these correlations as 
well, such as micrometeoroid bombardment, that would operate 
preferentially at smaller perihelion distances, where the particle 
number density might be higher. 

%_______________________________________________________________________________

\textbf{Dependences on ``sizes'':} We also searched for
possible dependences of the correlations on object ``sizes", which is equivalent to a dependence on absolute magnitudes ($H_{R}$) if one assumes the same albedo for the
objects. This should be a reasonable assumption because we analyze here only faint classical objects, which are expected to be a quite homogeneous sample. We obtain an average R-albedo ($p{_R}=0.12$) from 28 measured albedos of classical TNOs compiled from different authors (Brown \& Trujillo, 2004; Cruiksahnk et al. 2005; Grundy et al. 2005; Jewitt et al. 2001; \cite{Lykawka05}; \cite{Margot02}; \cite{Noll04}; \cite{Osip03}; \cite{Sheppard02}; \cite{Stansberry05}; \cite{Stansberry08}). One can speculate on such a dependence on sizes, since the 
smaller objects are probably collisionally more evolved. For this reason the 
largest objects should present different physico-chemical properties than 
the smallest ones. Peixinho et al. (2004) studied the color-perihelion trend 
for the classical objects, in particular for the B-R vs. q correlation. Using this cut-off brightness of $H_R < 6.2$mag as a proxy for a minimum size of the
objects (diameter D $>$ 190km for an albedo $p{_R}=0.12$), they find a 
significant increase of the correlation of the B-R color values with perihelion distance q 
for classical objects, and a decrease of the correlation when $H_{R}>6.2$ 
(D$<$190 km). Objects with shorter perihelion distance are closer to the Sun 
and would be susceptible to develop some sort of bound coma if they are large 
enough (D$>$150 km) (\cite{Del04}). Their surfaces may experience deposition
of fresh bluish ices when the object moves further away from perihelion, so
that the objects would appear bluer than before.
	
We find a slight increase in the B-R vs. q correlation for the classicals (OWN+LP+2MS) 
with $H_{R}<6.2$ (Table \ref{correlsizes1}), but with a not fully
convincing level of significance (SL). Similar results are obtained 
for B-V vs. q. Hence, we cannot conclude on firm color vs. 
perihelion distance dependence on size of the classical objects. 
Neither do we find dependence on sizes for the spectral gradients 
versus q. It is noteworthy, however, that opposite trends 
may exist for R-I vs. q in the sense that it increases for H$_{R}>6.2$, 
and decreases for H$_{R}<6.2$ (Table \ref{correlsizes1}). This result could imply that R-I correlates best with perihelion 
distance for the smaller classical objects, i.e. the smaller classical objects are bluer (less red) for the lower perihelion distances (q). This q-dependent coloring mechanism cannot be due to like-comet activity, because these objects are very small to retain a coma. Micrometeoroid bombardment or other q-dependence coloring mechanism might be necessary to explain this particular trend. This finding should be taken with
care since it effects only a single color and thus requires a very careful
physical interpretation that is currently not available. We do not find size 
dependencies versus aphelion distance (Q).

We find a possible dependence on sizes of the color correlations with 
semimajor axis (a), in particular for B-R vs. a (H$_{R}<6.2$) for the 
OWN+LP sample ($\rho=0.56$, SL=97.04\%). A similar trend is apparent 
for V-R vs. a (OWN+LP) and R-I vs. a (OWN+LP+2MS) as well. We reproduce this finding 
in the B-R vs. $\psi$ (H$_{R}<6.2$) correlation (Table \ref{correlsizes1}), 
however as a possible weak anti-correlation ($\rho$=-0.32, SL=92.06\%, n=32). These results are compatible since $\psi$ depends on $\frac{1}{a}$. They are also supported by our findings for Grt vs. $\psi$ (H$_{R}<6.2$), see Table \ref{Grt_correl}.

%_______________________________________________________________________________

Thus for the larger classical objects ($D>190 km$) there is an 
apparent trend of increased reddening with larger semi-major axis or, in other
words, they appear bluer the closer their mean distances to the Sun are. Based
on this finding, one may speculate that these bodies are large enough to retain 
volatiles produced either by impacts or intrinsic activity. Subsequent
deposition of the volatiles in icy form on the body surface could rejuvenate
its appearance with bluish (or less red) colors.

%_______________________________________________________________________________

%_______________________________________________________________________________

\subsection{Sub-populations among the classicals.} 

Tegler \& Romanishin (2000) found first indications that classical objects 
in near-circular orbits beyond 40 AU have very red colors and might thus 
constitute a separate group of objects in the Kuiper Belt. Other authors 
considered the existence of two separate groups as a function of 
orbital inclination (\cite{Brown01}; \cite{Levi01}; \cite{Doress}). 
The most reliable results about the existence of the two groups are 
shown in Peixinho et al. (2004). Based on their color-inclination 
properties and using a squared ranks test (\cite{Talw1977}; \cite{Cono1978}) 
Peixinho et al. obtained two distinct populations within the classical 
TNOs: the dynamically hot population (i$>4.5\,^{\circ}$) which show low
reddening values and a cluster of dynamically cold TNOs (i$<4.5\,^{\circ}$)
with very red colors. Hence, in the following we consider the possible 
correlations separately for the two populations, applying the selection
criterium by inclination.

%______________________________________________________________________

\subsubsection{Dynamically hot classicals (n=42)}
\label{hot}

The dynamically hot classicals are considered those with inclination $i > 4.5,^{\circ}$. The total sample (OWN+LP+2MS) contains 42 objects with colors 
measured of which our new observations contribute 24\%.

%_______________________________________________________________________________

\textbf{Strong correlations:} The main results obtained by other authors 
(\cite{Peix}; \cite{Doress_b}) are reproduced by us with minor 
differences. Particularly noteworthy are the strong correlations of colors versus i, and versus q 
(Table \ref{correl}) as for instance obtained also by Doressoundiram 
et al. (2005).

%_______________________________________________________________________________

\textbf{Weak correlations:} A weak (anti-)correlation is found  for B-V vs. e 
($\rho$=-0.34, SL=89.33\%, n=23). The spectral gradient of this sample, based
upon the B-V color values, varies from -2.5 to 41.8\%/100 nm with an 
average value of 19.3$\pm$ 6.2\%/100 nm. The anti-correlation found may
indicate that the redder hot classicals have the lower eccentricity orbits. 

%_______________________________________________________________________________

\begin{figure}
   \centering
   \includegraphics[angle=-90,width=9.0cm]{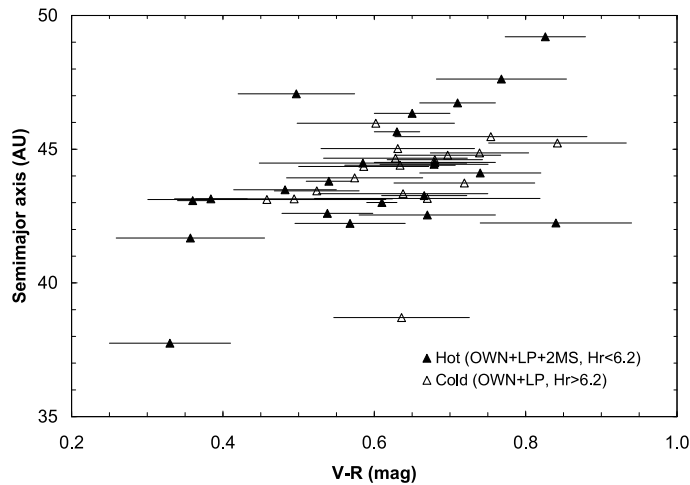}
   \caption{V-R vs. a for 22 (OWN+LP+2MS) hot classical objects (\emph{filled
triangles}) with $H_{R}<6.2$ ($D>190 km$) and for 17 (OWN+LP) cold classical
objects (\emph{open triangles}) with $H_{R}>6.2$ ($D<190 km$).}
         \label{VR_a_classicals}
   \end{figure}

\textbf{Dependence on ``sizes'':} We find the same trend for the dynamically hot population as for all classicals with regard to the color versus perihelion correlation (though with a slight increase in the SL when $H_{R}<6.2$ (D$>$190 km)), see Table \ref{correlsizes1}. Beyond that, we obtain a clear dependence on size for the color versus semimajor axis 
correlations (Table \ref{correlsizes1}), in particular for B-R vs. a ($H_{R}<6.2$, D$>$190 km). This correlation is complemented by those for V-R vs. $\psi$, Grt vs. a for H$_{R}<6.2$ (Tables \ref{correlsizes1}, and \ref{Grt_correl}) and V-R vs. a for $H_{R}<6.2$ (Fig. \ref{VR_a_classicals}, Table \ref{correlsizes1}). Similar correlations are
not found for the fainter objects ($H_{R}>6.2$). Finally, we would like to mention that
there might be a dependence on size for reddening gradients versus aphelion 
distance, as shown in Table \ref{Grt_correl} for Grt vs. Q ($H_{R}<6.2$). 
Size dependence is not seen  for the correlations of colors versus inclination and eccentricity, nor for spectral gradients.

%_______________________________________________________________________________

Thus, considering all the size dependences for the dynamically hot classicals 
(versus q, Q and a), we summarize that for the larger objects in this population (D$>190$ km) the redder ones have the larger perihelion \& aphelion distances, and semimajor axes. This is consistent with a resurfacing scenario by ice deposition due to a (temporal) bound coma as suggested by Delsanti et al. (2004). From our results we may be actually
seeing different behaviors for objects reflecting two different histories: resurfacing of objects with D$>190$ km might be dominated by the accretional history of the body, while that of objects with D$<190$ km might be dominated by their collisional history. However, our results are also compatible with the explanation scenario by Mc Kinnon (2002) for the blue coloring of the larger classical TNOs. He takes into account the internal evolution of the TNOs that are able to mobilize CO, e.g., through surface venting, and subsequent CO-ice deposition on the exterior, cold, porous layers. This outgassing behaviour could be accelerated in objects with smaller perihelion distances. Also, the TNO resurfacing by meteoroid bombardment may play a greater role for smaller distances, as mentioned in Section \ref{weaker_classicals}. At the moment, we cannot distinguish the importance of the various proposed scenarios for the population of dynamically hot classical TNOs. However, the various correlations found between surface and dynamical properties of this group support the existence of a separate population of dynamically hot classicals that are distinctly different from the other TNOs.  

%_______________________________________________________________________________

\subsection{Dynamically cold classicals (n=31)}
\label{cold}

This group of classicals (n=31) consists of objects with orbital inclination
$i < 4.5,^{\circ}$. Our new measurements of TNOs contribute 29\% of the total sample (OWN+LP+2MS) of dynamically cold classicals with colors measured. New and known (confirmed) correlations of color versus orbital parameters are discussed below.

%_______________________________________________________________________________

\textbf{Strong correlations:} We do not find strong correlations for the cold classicals objects (except a color-color one).

%_______________________________________________________________________________

\textbf{Weak correlations:} A new weak correlation is found for color R-I vs. 
q ($\rho$=0.27, SL=84.35\%, n=29; Table \ref{correlsizes1}). The highest colors vs. orbital 
parameter correlation has been obtained for V-R vs. a, which is a 
known weak correlation ($\rho$=0.35, SL=91.29\%, n=25; Table \ref{correlsizes1}). 
However, we believe that color photometry of more cold classicals is 
required to confirm these results.

%_______________________________________________________________________________

\textbf{Dependence on ``sizes'':} We find a stronger correlation for R-I vs. 
q ($\rho$=0.60, SL=98.06\%, n=16) for cold classicals with $H_{R}>6.2$, a
result that is obtained for the whole classical group as well 
(Table \ref{correlsizes1}). Since the hot classicals
do not show such a dependence, we conclude it is characteristic of the cold
population only. V-R vs. a correlates better for objects with $H_{R}>6.2$,
and worse for those with $H_{R}<6.2$ (Fig. \ref{VR_a_classicals}, 
Table \ref{correlsizes1}), i.e. the behaviour of the dynamically cold
classicals is opposite to that of the group of all classicals and in particular to that of the dynamically hot ones. Along with some distinct color-color correlations (B-V vs. V-R and V-R vs. R-I, on Table \ref{correl}) that display opposite spectral slopes 
for these two classical groups, it could be indicative of a different 
primordial origin of the hot and cold classical populations. According 
to Morbidelli et al. (2003), the \emph{dynamically hot bluish} population 
could have formed in a region closer to the Solar System than the 
\emph{dynamically cold reddish} one. Neptune's migration is made responsible
for shifting both populations to their current distance range in the 
Kuiper Belt. Based upon the correlations discussed above, one can speculate that 
a coloring mechanism may exist that acts in the opposite way and depending on
q and a for the smaller cold classicals, compared to the hot ones.

%_______________________________________________________________________________

%_______________________________________________________________________________

\subsection{Scattered disk objects SDOs (n=25)}
\label{SDOs}

The total sample of SDOs with colors measured comprises 25 objects. Our new
measurements contribute 40\%\footnote{Including the peculiar distant object 2000CR$_{105}$.} to the total sample (OWN+LP+2MS). Our correlation
analysis has led to results that are very similar to the ones obtained 
by other authors (\cite{Peix}; \cite{Doress_b}). 

%_______________________________________________________________________________

\begin{figure}
   \centering
   \includegraphics[angle=-90,width=9.0cm]{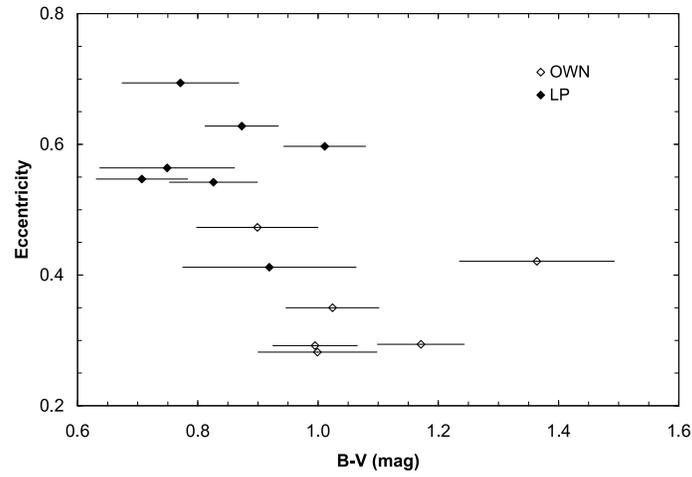}
      \caption{B-V vs. e for 13 SDOs (OWN+LP). Symbols: OWN sample
(\emph{open diamonds}) and LP sample (\emph{filled diamonds}).}
         \label{BV_e_SDOs}
   \end{figure}

\textbf{Strong correlations:} Besides correlations versus q, H$_{R}$, and e as
mentioned  by Peixinho et al. (2004) and Doressoundiram et al. (2005), we
find correlations for B-V vs. q, B-V vs. H$_{R}$, B-V vs. e, and R-I
vs. e (for the correlation characteristics see Table \ref{correl}). Contrary
to Doressoundiram et al. (2005) who claim a weak correlation for B-R vs. e, 
we find an anti-correlation for B-V vs. e. This finding (Fig. \ref{BV_e_SDOs})
is supported by the correlation results for Grt vs. e (Table \ref{Grt_correl}) 
and could mean that SDOs with lower eccentricities are redder than the higher
eccentricity ones. In consequence, the SDOs in high eccentricity orbits may
be more subject to a resurfacing mechanism that produces bluish colors. From the
known correlations B-R vs. H$_{R}$ displays the highest confidence level 
($\rho$=-0.81, SL=99.49\%, n=13). Overall, the correlation results suggest
that color vs. H$_{R}$ seem to play an important role in the SDOs group 
(Table \ref{correl}). This is also seen for the respective Grt vs H$_{R}$
correlations (see Table \ref{Grt_correl}). We find 
a similar trend as found by Peixinho et al. (2004) for B-R vs. q, although the
significance level seems to decrease for larger sample sizes (n=13: 
$\rho$=0.62, SL=96.85\%; n=19: $\rho$=0.34, SL=85.54\%; see Table \ref{correl}). 

%_______________________________________________________________________________

\textbf{Weak correlations:} Our analysis provides evidence for a weaker 
correlation between B-V and i that is known from previous datasets 
(Table \ref{correl}). This correlation could imply that SDOs in higher 
inclination orbits are redder than those with lower inclination. Since the
sample size is small (n=13), additional SDO photometry would allow us to confirm or disprove the weak correlation mentioned above. 

We found some additional weak correlations between colors and orbital elements as follows: R-I vs. q and B-R vs. e.

Regarding H$_{R}$, inclination and eccentricity correlations we can say 
that the SDOs with fainter absolute magnitudes (i.e. the smaller ones), 
low inclinations, and high eccentricities appear typically
bluer. Unfortunately, the abovementioned resurfacing scenarios are unable to
account for these (weak) correlations and no simple mechanism is known 
that could produce such coloring profiles properly. So, one may have to think
of combined effects, and in any case reevaluate the respective
correlation characteristics based upon a better and larger SDO sample.

%_______________________________________________________________________________

\subsection{Combined groups}

In order to analyse possible color relationships among the various dynamical
groups of TNOs we consider in the following two samples of combined
population, e.g. SDOs with hot classicals and SDOs with Centaurs. The common
property of the combined groups is that their orbits were subject to
gravitational scattering in the past. Hence, by comparing the results of the
combined samples with those of the individual ones described above, it may be
possible to isolate possible similarities and diversities that may allow us to
constrain common or different origin and/or evolution. For the analysis, one
must take into account that in the combined sample, individual groups have
different weights according to the respective sample sizes. 

%_______________________________________________________________________________

\subsubsection{Merged sample: SDOs+hot classicals (n=67)}

The total number of objects in this sample is 67 of which 42 are hot
classicals and 25 are SDOs.

Some correlation evidence in the merged sample increases with respect 
to the SDOs alone, for instance V-R vs. i, and B-R vs. i. There are also some that disappear, for instance B-V vs. H$_{R}$, B-V vs. e, V-R vs. H$_{R}$, R-I vs. q, R-I vs. e, B-R vs. H$_{R}$, and B-R vs. e. The increase in the color vs. i correlations 
is clearly due to the hot classicals in the merged sample (Table \ref{correl}). For the comparison between the merged sample and hot classicals we again have a similar situation. Here, we would like to mention in particular the less prominent or vanishing correlations, e.g. B-V vs. i, B-V vs. e, V-R vs. a, B-R vs. q, B-R vs. i, B-R vs. e, and B-R vs. a (Table \ref{correl}).

In summary, there is no significant strengthening for color-orbit correlations 
seen in the merged group of SDOs and hot classicals such that we can argue for
a common origin or very similar evolution.

%_______________________________________________________________________________

\subsubsection{Merged sample: SDOs+Centaurs (n=43)}

The merged sample of 43 objects consists of 25 SDOs and 18 Centaurs.

The situation for the SDOs+Centaurs sample compared to SDOs and Centaurs alone
is again unchanged, with a few moderately stronger correlations compared to
SDOs alone on one side, (for instance for R-I vs. i, and B-R vs. a) and weaker
or disappearing correlations on the other side (B-V vs. q, B-V vs. H$_{R}$,
B-V vs. i, B-V vs. e, V-R vs. q, V-R vs. H$_{R}$, R-I vs. q, R-I vs. e, B-R
vs. q, B-R vs. H$_{R}$, and B-R vs. e). The increase of the correlation
evidence in color vs. i and a is possibly a contribution from the 
Centaur group. Compared to the Centaurs alone we find increased evidence for
correlations between R-I and e, B-R and q, and B-R and H$_{R}$, certainly due
to the SDO contribution in the merged sample. On the other side decreased or
disappearing evidence is found in the merged sample for the correlations 
B-V vs. i, B-V vs. e, V-R vs. q, V-R vs. H$_{R}$, V-R vs. e, V-R vs. a, 
R-I vs. H$_{R}$, R-I vs. i, B-R vs. i, B-R vs. e, and B-R vs. a.

In summary, as for the other merged sample, no clear changes in the correlation significances are identified that would allow us to argue for or against a possible common origin and/or evolution of the SDOs and Centaurs.

%_______________________________________________________________________________

\section{Conclusions}

   \begin{itemize}

      \item BVRI photometry results on 32 KBOs with ESO/VLT telescope are 
presented (colors and absolute magnitudes). Merging our photometric results 
with the ESO and 2MS data sets, we obtain a large and rather homogeneous
sample of 116 objects that is used for statistical analysis of correlations
between colors and/or orbital parameters of the objects.

	  \item We found short-term brightness variability due to rotation 
in 11 of the 32 objects studied. While the percentage of objects with very
high amplitudes is similar to results published before, our sample has a
higher fraction of objects with medium to small amplitude variations. Since 
our sample includes objects with fainter magnitudes (i.e. smaller sizes) 
than previous studies, we tentatively conclude that smaller objects have, on average, more irregular shapes than larger ones. Independent confirmation of this conclusion with a much larger sample is highly desirable.

	  \item Our average spectral gradient calculations are compatible 
with previous results:  

	   Grt$_{avg}$(Class, n=73)=22.9 $\pm$ 9.8\%/100 nm
	   
	   Grt$_{avg}$(Hot, n=42)=19.6 $\pm$ 8.5\%/100 nm
	   
	   Grt$_{avg}$(Cold, n=31)=27.4 $\pm$ 11.3\%/100 nm
	   
	   Grt$_{avg}$(SDOs, n=25)=18.6 $\pm$ 7.6\%/100 nm
	   
	   Grt$_{avg}$(Cent, n=18)=19.8 $\pm$ 6.7\%/100 nm

	  \item We obtain color vs. perihelion distance correlations, and color vs. inclination 
and eccentricity anticorrelations for the classical TNOs. These
results imply that the redder classical objects have low inclinations and
eccentricities, and high perihelion distances. This interpretation supports
the existence of a \emph{cold red cluster} of classical objects in very
circular orbits with low inclinations, and possibly supports an ice
resurfacing scenario that depends on the solar distance.	  

	  \item A size dependence of the correlation of colors versus semimajor axis is found in classical objects. We use an absolute magnitude cuttoff value of H$_{R}$=6.2.  For the large objects (H$_{R}<$6.2, D$>$190 km, assuming p$_{R}$=0.12) the redder ones have the larger semimajor axes. This result could support a possible resurfacing mechanism depending on semimajor axis. The large objects sizes might allow them to develop a bound coma, and perhaps a coloring mechanism due to sublimation, and recondensation can take place. No apparent dependence on sizes of the correlations of colors versus perihelion distances, inclination or eccentricity are found for classical objects.	  

	  \item We find in the hot (i$>4.5\,^{\circ}$) classical objects
that the redder ones have the larger perihelion distances, and the lower orbital inclinations. We find that correlations of color versus perihelion distance, aphelion distance and semimajor axis depend on size: these correlations increase for the larger (D$>190$ km) hot classical objects. These results support a coloring mechanism that depends on solar distance and size.	 		  

	  \item Two possible color versus semimajor axis, and perihelion distance correlations are found for the cold (i$<4.5\,^{\circ}$) group. These results seem to have a relevant dependence on sizes: the smaller (D$<190$ km) cold classical objects correlate better
with perihelion and semimajor axis than the bigger ones. In other words, the smaller redder cold objects would have higher semimajor axe and perihelion distances. So, it seems there is a solar distance dependent coloring mechanism different to ice resurfacing from sublimation and redeposition for the smaller cold objects (the small mass does not permit the development of a bound coma).

	  \item A possible different primordial origin for the hot and cold populations is consistent with our correlations and statistical analysis. We find seven higher correlations for the hot group than for the classical one. The V-R vs. a correlation presents an opposite trend in size for the hot and cold groups: for the hot objects the V-R vs. semimajor axis increases for the larger ones (H$_{R}<$6.2, D$>$190 km), and for the cold objects it increases for the smaller ones (H$_{R}>$6.2, D$<$190 km).	  

	  \item The SDOs with higher absolute magnitudes (i.e. the smaller ones), lower inclinations and perihelion distances, and higher eccentricities are the bluer ones. A possible physical process that may produce such a coloring scheme is currently unknown to us.
	  
	  \item The Centaurs with higher inclinations, lower eccentricities, smaller sizes and lower semimajor axis are the bluer ones. This result is opposite to that obtained for the SDOs, so the coloring mechanism for Centaurs and SDOs must be different.

	  \item We do not find relevant differences between the statistical results for merged groups (i.e. SDOs+hot, SDOs+Cent), so this cannot be used to link different dynamical groups or to say something about a possible common origin and/or evolutionary relation between these different dynamical groups.	  	

   \end{itemize}
   
%_______________________________________________________________________________

\begin{acknowledgements}

We are grateful to the Cerro Paranal Observatory staff. This research was based on data obtained at the VLT on Cerro Paranal Observatory, which is operated by the European Southern Observatory (ESO). This work was supported by contracts AYA-2002-00382, AYA-2004-03250, and AYA-2008-06202-C03-01. P.S.S. wishes to acknowledge J.D. Santander-Vela for his help with the manuscript.
      
\end{acknowledgements}

%_______________________________________________________________________________

\clearpage

%_______________________________________________________________________________

%\clearpage
%_______________________________________________________________________________
\begin{table*}

\caption{Object sample and observational circumstances} % title of Table

\label{tableOBS} % is used to refer this table in the text

\centering % used for centering table

\begin{tabular}{c c c c c c c c} % centered columns (4 columns)

\hline\hline % inserts double horizontal lines

& Object & Type & UT Date &  $\Delta$(AU) & r (AU) & $\alpha$ (deg) & Opps.\\
% table heading

\hline % inserts single horizontal line

&1997SZ$_{10}$  & 2:1 & 2003 Nov 2  & 31.139 & 32.104 & 0.4 & 3\\
% inserting body of the table

&2000CJ$_{105}$ & Classical(hot) & 2004 Mar 26 & 46.893 & 47.829 & 0.4 & 4\\

(60458)&2000CM$_{114}$  & SDO & 2004 Mar 26 & 42.100 & 42.759 & 1.0 & 4\\

&2000CN$_{105}$ & Classical(cold) & 2004 Mar 26 & 45.233 & 46.074 & 0.7 & 4\\

&2000CP$_{104}$ & Classical(hot) & 2004 Mar 25 & 45.651 & 46.499 & 0.7 & 3\\

&2000CR$_{105}$ & ESDO & 2004 Mar 26 & 53.930 & 54.612 & 0.8 & 3\\

&2000YB$_{2}$ & Classical(cold) & 2003 Nov 2 & 37.616 & 38.535 & 0.6 & 3\\

&2000YC$_{2}$ & SDO & 2003 Nov 2 & 38.617 & 39.540 & 0.5 & 3\\

&2000YU$_{1}$ & Classical(hot) & 2003 Nov 1 & 43.290 & 44.218 & 0.5 & 3\\

(82075)&2000YW$_{134}$ & SDO & 2004 Mar 25 & 42.750 & 43.190 & 1.2 & 4\\

&2001FM$_{194}$ & SDO & 2004 Mar 26 & 35.439 & 36.435 & 0.1 & 3\\

&2001HY$_{65}$ & Classical(hot) & 2004 Mar 25 & 38.305 & 39.299 & 0.1 & 4\\

&2001HZ$_{58}$ & Classical(cold) & 2004 Mar 25 & 42.460 & 43.269 & 0.8 & 2\\

&2001QC$_{298}$ & Classical(hot) & 2003 Nov 1 & 40.051 & 40.563 & 1.2 & 4\\

&2001QO$_{297}$ & Classical(cold) & 2003 Nov 1 & 42.724 & 43.386 & 1.0 & 3\\

&2001QP$_{297}$ & Classical(cold) & 2003 Nov 1 & 42.810 & 43.522 & 0.9 & 3\\

&2001QX$_{322}$ & SDO & 2003 Nov 2 & 38.599 & 39.387 & 0.9 & 2\\

&2001RZ$_{143}$ & Classical(cold) & 2003 Nov 2 & 40.458 & 41.402 & 0.4 & 3\\

&2001SQ$_{73}$ & Cent & 2003 Nov 2 & 15.021 & 15.999 & 0.6 & 3\\

(42301)&2001UR$_{163}$ & SDO & 2003 Nov 2 & 48.313 & 49.264 & 0.3 & 6\\

&2002CC$_{249}$ & Classical(cold) & 2004 Mar 25 & 37.743 & 38.699 & 0.4 & 3\\

&2002CX$_{154}$ & SDO & 2004 Mar 25 & 37.684 & 38.629 & 0.5 & 3\\

&2002CY$_{224}$ & SDO & 2004 Mar 26 & 35.589 & 36.208 & 1.2 & 3\\

&2002GH$_{32}$ & Classical(hot) & 2004 Mar 25 & 42.009 & 42.848 & 0.7 & 3\\

&2002GJ$_{32}$ & Classical(hot) & 2004 Mar 26 & 42.132 & 42.952 & 0.8 & 3\\

(84522)&2002TC$_{302}$ & SDO & 2003 Nov 2 & 47.080 & 48.055 & 0.2 & 4\\

(55637)&2002UX$_{25}$ & Classical(hot) & 2003 Nov 1 & 41.575 & 42.540 & 0.3 &
5\\

&2002VT$_{130}$ & Classical(cold) & 2003 Nov 2 & 41.768 & 42.688 & 0.5 & 2\\

&2003AZ$_{84}$ & 3:2(Plutino)& 2003 Nov 2 & 45.492 & 45.844 & 1.2 & 4\\

&2003QB$_{112}$& Classical(hot) & 2003 Nov 1 & 38.972 & 39.734 & 0.9 &
(57d.)\\

&2003QW$_{90}$ & Classical(hot) & 2003 Nov 1 & 43.718 & 44.484 & 0.8 & 2\\

&2003QY$_{111}$& Classical(cold) & 2003 Nov 1 & 41.359 & 42.088 & 0.9 & 2\\

\hline %inserts single line

\end{tabular}

\footnotesize{\emph{Type:} Dynamical classification of the objects as described in section \ref{DataAnalysis}. \emph{$\Delta$(AU):} Distance Object-Earth in Astronomical Units. \emph{r
(AU):} Distance Object-Sun in Astronomical Units. \emph{$\alpha$ (deg):}
Phase angle in degrees. \emph{Opps.}, is the number of observed oppositions
of the object for the observation date.}

\end{table*}

\begin{table*}
\caption{Photometric measurements of our TNO sample} % title of Table
\label{tableRES} % is used to refer this table in the text
\centering % used for centering table
\begin{tabular}{c c c c c c c c c} % centered columns (4 columns)
\hline\hline % inserts double horizontal lines
Object & R & B-V & V-R & R-I & V-I & B-R & B-I & Grt \\ % table heading
       & (mag)&(mag)&(mag)&(mag)&(mag)&(mag)&(mag)&(\%$/100 nm$)\\

\hline % inserts single horizontal line 
Solar colors & & 0.67 & 0.36 & 0.33 & 0.69 & 1.03 & 1.36\\
\hline

1997SZ$_{10}$   & 23.01$\pm$0.09 &             & 0.85$\pm$0.12 & 0.57$\pm$0.10 & 1.42$\pm$0.09 &   	 		 & 	 & 43.1$\pm$18.4\\ % inserting body of the table
2000CJ$_{105}$  & 22.44$\pm$0.04 & 1.08$\pm$0.08 & 0.68$\pm$0.07 & 0.70$\pm$0.06 & 1.38$\pm$0.07 & 1.76$\pm$0.07 & 2.46$\pm$0.06&  33.3$\pm$8.1\\
(60458)2000CM$_{114}$& 23.21$\pm$0.04 & 			 & 			   & 0.66$\pm$0.09 & 0.75$\pm$0.09 & 			 & 	& 16.0$\pm$5.0\\
2000CN$_{105}$	& 21.72$\pm$0.06 & 1.06$\pm$0.10 & 0.66$\pm$0.08 & 0.69$\pm$0.08 & 1.35$\pm$0.08 & 1.72$\pm$0.10 & 2.41$\pm$0.10& 32.0$\pm$10.4\\
2000CP$_{104}$	& 23.22$\pm$0.03 & 			 & 0.53$\pm$0.06 & 0.64$\pm$0.06 & 1.17$\pm$0.07 & 			 & & 21.4$\pm$7.3\\
2000CR$_{105}$	& 23.47$\pm$0.06 & 			 & 0.28$\pm$0.08 & 0.87$\pm$0.09 & 1.16$\pm$0.09 & 			 &	 & 15.3$\pm$9.9\\
2000YB$_{2}$	& 22.24$\pm$0.06 & 0.86$\pm$0.10 & 0.64$\pm$0.09 & 0.55$\pm$0.07 & 1.18$\pm$0.08 & 1.49$\pm$0.09 & 2.04$\pm$0.08&  21.2$\pm$9.2\\
2000YC$_{2}$	& 23.04$\pm$0.05 & 			 & 0.67$\pm$0.11 & 0.24$\pm$0.11 & 0.91$\pm$0.10 & 			 & 	& 14.2$\pm$12.5\\
2000YU$_{1}$	& 22.83$\pm$0.03 & 			 & 0.66$\pm$0.06 & 0.52$\pm$0.04 & 1.19$\pm$0.06 &			 &	&  25.5$\pm$6.7\\
(82075)2000YW$_{134}$& 20.71$\pm$0.06 & 1.00$\pm$0.07 & 0.45$\pm$0.08 & 0.52$\pm$0.08 & 0.97$\pm$0.08 &			 &	&  15.4$\pm$7.4\\
2001FM$_{194}$	& 22.83$\pm$0.04 & 			 & 0.47$\pm$0.06 & 0.62$\pm$0.06 & 1.09$\pm$0.07 &			 &	&  16.2$\pm$6.9\\
2001HY$_{65}$	& 21.93$\pm$0.05 & 1.03$\pm$0.06 & 0.48$\pm$0.07 & 0.71$\pm$0.06 & 1.19$\pm$0.05 & 1.51$\pm$0.06 & 2.22$\pm$0.05&  22.6$\pm$6.6\\
2001HZ$_{58}$	& 22.48$\pm$0.05 & 1.17$\pm$0.06 & 0.47$\pm$0.06 & 0.70$\pm$0.06 & 1.17$\pm$0.05 & 1.64$\pm$0.06 & 2.34$\pm$0.05 &  24.5$\pm$6.1\\
2001QC$_{298}$	& 22.51$\pm$0.05 & 0.66$\pm$0.07 & 0.37$\pm$0.07 & 0.63$\pm$0.07 & 1.00$\pm$0.06 & 1.03$\pm$0.08 & 1.66$\pm$0.06 &  6.9$\pm$6.7\\
2001QO$_{297}$	& 22.31$\pm$0.09 & 			 & 0.69$\pm$0.11 & 0.58$\pm$0.10 & 1.27$\pm$0.07 & 			 & 	&  30.1$\pm$14.6\\
2001QP$_{297}$  & 23.03$\pm$0.07 &        	 & 0.63$\pm$0.10 & 0.60$\pm$0.09 & 1.23$\pm$0.10 &			 &	&  26.6$\pm$12.5\\
2001QX$_{322}$	& 21.99$\pm$0.06 & 1.36$\pm$0.13 & 0.65$\pm$0.11 & 0.42$\pm$0.09 & 1.07$\pm$0.11 & 2.01$\pm$0.11 & 2.43$\pm$0.11 &  27.7$\pm$10.4\\
2001RZ$_{143}$	& 22.36$\pm$0.08 & 1.08$\pm$0.14 & 0.51$\pm$0.13 & 0.49$\pm$0.10 & 1.00$\pm$0.12 & 1.59$\pm$0.12 & 2.07$\pm$0.12 &  19.0$\pm$11.5\\
2001SQ$_{73}$	& 20.68$\pm$0.06 & 0.52$\pm$0.09 & 0.50$\pm$0.09 & 0.57$\pm$0.09 & 1.07$\pm$0.09 & 1.02$\pm$0.08 & 1.59$\pm$0.08 &  6.2$\pm$9.6\\
(42301)2001UR$_{163}$& 20.41$\pm$0.06 & 1.00$\pm$0.10 & 1.06$\pm$0.08 & 0.91$\pm$0.07 & 1.96$\pm$0.07 & 2.05$\pm$0.10 & 2.96$\pm$0.09 &  68.1$\pm$16.1\\
2002CC$_{249}$	& 21.87$\pm$0.05 & 			 & 0.51$\pm$0.08 & 0.69$\pm$0.06 & 1.20$\pm$0.07 & 			 & 	&  22.3$\pm$8.5\\
2002CX$_{154}$	& 22.99$\pm$0.05 & 0.90$\pm$0.10 & 0.57$\pm$0.08 & 0.50$\pm$0.07 & 1.07$\pm$0.08 & 1.47$\pm$0.09 & 1.97$\pm$0.09 &  17.6$\pm$8.2\\
2002CY$_{224}$	& 21.42$\pm$0.04 & 1.02$\pm$0.08 & 0.66$\pm$0.06 & 0.64$\pm$0.06 & 1.30$\pm$0.06 & 1.69$\pm$0.07 & 2.33$\pm$0.07 &  29.2$\pm$7.1\\
2002GH$_{32}$	& 22.35$\pm$0.05 & 0.99$\pm$0.07 & 0.57$\pm$0.07 & 0.59$\pm$0.07 & 1.16$\pm$0.06 & 1.56$\pm$0.07 & 2.15$\pm$0.06 &  22.3$\pm$7.4\\
2002GJ$_{32}$	& 21.69$\pm$0.09 & 1.33$\pm$0.12 & 0.59$\pm$0.14 & 0.42$\pm$0.12 & 1.01$\pm$0.13 & 1.92$\pm$0.11 & 2.34$\pm$0.10 &  24.4$\pm$12.7\\
(84522)2002TC$_{302}$& 20.27$\pm$0.06 & 1.17$\pm$0.07 & 0.58$\pm$0.08 & 0.67$\pm$0.07 & 1.25$\pm$0.06 & 1.75$\pm$0.07 & 2.42$\pm$0.06 &  28.9$\pm$7.9\\
(55637)2002UX$_{25}$& 19.64$\pm$0.04 & 0.94$\pm$0.06 & 0.54$\pm$0.06 & 0.59$\pm$0.05 & 1.13$\pm$0.05 & 1.48$\pm$0.06 & 2.07$\pm$0.05 &  19.8$\pm$5.8\\
2002VT$_{130}$	& 21.68$\pm$0.07 & 1.45$\pm$0.21 & 0.56$\pm$0.10 & 0.63$\pm$0.09 & 1.19$\pm$0.10 & 2.00$\pm$0.20 & 2.63$\pm$0.20 &  30.8$\pm$11.0\\
2003AZ$_{84}$	& 20.14$\pm$0.04 & 0.61$\pm$0.08 & 0.45$\pm$0.07 & 0.35$\pm$0.06 & 0.80$\pm$0.07 & 1.06$\pm$0.07 & 1.41$\pm$0.07 &  1.6$\pm$6.4\\
2003QB$_{112}$	& 22.63$\pm$0.04 & 			 & 0.75$\pm$0.07 & 0.01$\pm$0.06 & 0.77$\pm$0.07 & 			 & 		&  12.0$\pm$7.1\\
2003QW$_{90}$	& 21.17$\pm$0.03 & 1.11$\pm$0.07 & 0.67$\pm$0.06 & 0.65$\pm$0.05 & 1.32$\pm$0.06 & 1.78$\pm$0.06 & 2.43$\pm$0.06 &  31.5$\pm$6.2\\
2003QY$_{111}$	& 23.48$\pm$0.06 & 			 & 0.46$\pm$0.16 & 0.69$\pm$0.07 & 1.15$\pm$0.15 & 			 & 		&  18.5$\pm$15.5\\
\hline %inserts single line
\end{tabular}
\end{table*}

\begin{table*}
\caption{Objects with possible short-term brightness variability} % title of Table

\label{BrightVAR} % is used to refer this table in the text
\centering % used for centering table
\begin{tabular}{c c c c c c c} % centered columns (4 columns)
\hline\hline % inserts double horizontal lines
  & Object & Type & $\Delta$t(min.) & $\Delta$R(mags.) & $\sigma_\mathrm{phot}$ & R-range\\
\hline % inserts single horizontal line 
3$\sigma$ var. 		& 1997SZ$_{10}$ & 2:1  & 33	 & 0.36 & 0.06 & 22.82-23.18\\
							& 2000CR$_{105}$& ESDO & 50  & 0.21 & 0.05 & 23.28-23.49\\
							& 2000YC$_{2}$  & SDO  & 42  & 0.33 & 0.10 & 22.86-23.19\\						
							& 2000YU$_{1}$	& Hot  & 44	 & 0.12 & 0.03 & 23.77-23.89\\
							& 2003AZ$_{84}$ & 3:2  & 18  & 0.20 & 0.01 & 20.04-20.24\\
							& 2003QB$_{112}$& Hot  & 151 & 0.48 & 0.05 & 22.40-22.88\\
							& 2003QY$_{111}$& Cold & 156 & 0.72 & 0.11 & 23.10-23.82\\
$\leq 2\sigma$ var.		    & 2000CP$_{104}$& Hot  & 58  & 0.08 & 0.04 & 23.17-23.25\\
							& 2001QX$_{322}$& SDO  & 24  & -- & 0.05 & 21.97-22.01\\ 
							& 2002UX$_{25}$ & Hot  & 46  & -- & 0.01 & 19.63-19.64\\	
							& 2003QW$_{90}$ & Hot  & 18  & -- & 0.01 & 21.16-21.19\\						 
\hline %inserts single line
\end{tabular}

\footnotesize{\emph{Type}, is the dynamical classification of the objects as described in section \ref{DataAnalysis}:
hot (classical with i$>4.5\,^{\circ}$), cold (classical with i$<4.5\,^{\circ}$), SDO (scattered disk object), ESDO (extended scattered disk object), 2:1 and 3:2 (Neptune{'}s resonant objects). \emph{$\Delta$t}, is the measurement interval in minutes. \emph{$\Delta$R}, is the R-magnitude variation. $\sigma_\mathrm{phot}$ is the uncertainty of the relative photometry of the object (see text). R-range, is the total range of the R magnitudes measured. We do not include the $\Delta$Rs that are less than $\sigma_\mathrm{phot}$.}
\end{table*}

\begin{table*}
\caption{Absolute magnitudes for the TNOs of our sample} % title of Table
\label{AbsMag} % is used to refer this table in the text
\centering % used for centering table
\begin{tabular}{c c c c c} % centered columns (4 columns)
\hline\hline % inserts double horizontal lines
Object & $H_{V}(Bowell)$ & $H_{V}(Linear)$ & $H_{R}(Bowell)$ & $H_{R}(Linear)$\\
\hline % inserts single horizontal line 
1997SZ$_{10}$ &	8.75$\pm$0.08	& 8.80$\pm$0.08 &	7.90$\pm$0.09	& 7.95$\pm$0.09\\
2000CJ$_{105}$&	6.26$\pm$0.06	& 6.31$\pm$0.06 &	5.58$\pm$0.05	& 5.63$\pm$0.04\\
2000CM$_{114}$&	6.82$\pm$0.05	& 6.87$\pm$0.05 &	6.74$\pm$0.04	& 6.78$\pm$0.04\\
2000CN$_{105}$&	5.63$\pm$0.06	& 5.68$\pm$0.06 &	4.97$\pm$0.07	& 5.02$\pm$0.06\\
2000CP$_{104}$&	6.96$\pm$0.06	& 7.01$\pm$0.06 &	6.43$\pm$0.03	& 6.48$\pm$0.03\\
2000CR$_{105}$&	6.24$\pm$0.05	& 6.28$\pm$0.05 &	5.96$\pm$0.06	& 6.00$\pm$0.06\\
2000YB$_{2}$  &	6.93$\pm$0.07	& 6.98$\pm$0.07 &	6.30$\pm$0.06	& 6.34$\pm$0.06\\
2000YC$_{2}$  &	7.82$\pm$0.08	& 7.86$\pm$0.07 &	7.00$\pm$0.05	& 7.05$\pm$0.05\\
2000YU$_{1}$  &	6.97$\pm$0.05	& 7.01$\pm$0.05 &	6.30$\pm$0.03	& 6.35$\pm$0.03\\
2000YW$_{134}$&	4.61$\pm$0.06	& 4.65$\pm$0.06 &	4.16$\pm$0.06	& 4.20$\pm$0.06\\
2001FM$_{194}$&	7.70$\pm$0.05	& 7.73$\pm$0.05 &	7.23$\pm$0.04	& 7.26$\pm$0.04\\
2001HY$_{65}$ &	6.48$\pm$0.05	& 6.51$\pm$0.05 &	6.00$\pm$0.05	& 6.03$\pm$0.05\\
2001HZ$_{58}$ &	6.47$\pm$0.05   & 6.51$\pm$0.04 &	5.99$\pm$0.05	& 6.04$\pm$0.05\\
2001QC$_{298}$&	6.62$\pm$0.05	& 6.65$\pm$0.05 &	6.25$\pm$0.06	& 6.28$\pm$0.05\\
2001QO$_{297}$&	6.47$\pm$0.07	& 6.51$\pm$0.06 &	5.78$\pm$0.10	& 5.82$\pm$0.09\\
2001QP$_{297}$&	7.14$\pm$0.08	& 7.18$\pm$0.08 &	6.51$\pm$0.07	& 6.55$\pm$0.07\\
2001QX$_{322}$&	6.55$\pm$0.10	& 6.60$\pm$0.09 &	5.91$\pm$0.06	& 5.95$\pm$0.06\\
2001RZ$_{143}$&	6.65$\pm$0.10	& 6.69$\pm$0.10 &	6.14$\pm$0.08	& 6.18$\pm$0.08\\
2001SQ$_{73}$ &	9.14$\pm$0.07	& 9.21$\pm$0.07 &	8.64$\pm$0.07	& 8.71$\pm$0.06\\
2001UR$_{163}$&	4.49$\pm$0.06	& 4.53$\pm$0.06 &	3.44$\pm$0.06	& 3.48$\pm$0.06\\
2002CC$_{249}$&	6.46$\pm$0.06	& 6.50$\pm$0.06 &	5.95$\pm$0.05	& 5.99$\pm$0.05\\
2002CX$_{154}$&	7.62$\pm$0.07	& 7.67$\pm$0.07 &	7.06$\pm$0.05	& 7.10$\pm$0.05\\
2002CY$_{224}$&	6.32$\pm$0.05	& 6.35$\pm$0.05 &	5.66$\pm$0.04	& 5.69$\pm$0.04\\
2002GH$_{32}$ &	6.49$\pm$0.05	& 6.54$\pm$0.05 &	5.92$\pm$0.06	& 5.97$\pm$0.05\\
2002GJ$_{32}$ &	5.83$\pm$0.11	& 5.87$\pm$0.11 &	5.24$\pm$0.09	& 5.28$\pm$0.09\\
2002TC$_{302}$&	4.01$\pm$0.05	& 4.05$\pm$0.05 &	3.43$\pm$0.06	& 3.47$\pm$0.06\\
2002UX$_{25}$ &	3.85$\pm$0.05	& 3.89$\pm$0.04 &	3.31$\pm$0.04	& 3.36$\pm$0.04\\
2002VT$_{130}$&	5.87$\pm$0.08	& 5.91$\pm$0.08 &	5.31$\pm$0.07	& 5.35$\pm$0.07\\
2003AZ$_{84}$ &	3.78$\pm$0.06	& 3.81$\pm$0.05 &	3.33$\pm$0.05	& 3.36$\pm$0.04\\
2003QB$_{112}$&	7.26$\pm$0.06	& 7.30$\pm$0.05 &	6.50$\pm$0.04	& 6.54$\pm$0.04\\
2003QW$_{90}$ &	5.23$\pm$0.05	& 5.27$\pm$0.05 &	4.56$\pm$0.03	& 4.61$\pm$0.03\\
2003QY$_{111}$&	7.56$\pm$0.15	& 7.60$\pm$0.15 &	7.10$\pm$0.07	& 7.14$\pm$0.06\\ 
\hline %inserts single line
\end{tabular}

\footnotesize{V and R absolute magnitudes for the observed objects estimated
using the Bowell formalism ($H_{V}(Bowell)$, $H_{R}(Bowell)$), and the linear
approximation ($H_{V}(Linear)$, $H_{R}(Linear)$), as is explained in the
text.}
\end{table*}

%\clearpage

\begin{table}
\caption{Average spectral gradients for TNOs and Centaurs} % title of Table
\label{Avg_Grt} % is used to refer this table in the text
\centering % used for centering table
\begin{tabular}{c c c} % centered columns (4 columns)
\hline\hline % inserts double horizontal lines
& Grt & n\\
& (\%$/100 nm$)& \\
  \hline % inserts single horizontal line 
Classical & 22.9$\pm$9.8 & 73\\ % inserting body of the table
Hot	    & 19.6$\pm$ 8.5 & 42\\
Cold	& 27.4$\pm$11.3 & 31\\	
\hline
SDOs	& 18.6$\pm$ 7.6 & 25\\
\hline
Centaurs    & 19.8$\pm$ 6.7 & 18\\
\hline %inserts single line
\end{tabular}

\footnotesize{Average spectral gradients for the different dynamical 
groups (except resonant objects) calculated for the whole sample: OWN+LP+2MS. \emph{Grt}, is the 
spectral gradient calculated as explained in Section \ref{Grt}. The 
errors give the statistical error of the average color and fitted 
spectral gradient values, respectively. \emph{n}, is the total 
number of objects used in the average calculation.}
\end{table}

\begin{table}
\caption{Object summary} % title of Table
\label{Summary} % is used to refer this table in the text
\centering % used for centering table
\begin{tabular}{c c c c c c c} % centered columns (4 columns)
\hline\hline % inserts double horizontal lines
  & & & Types & & & \\ % table heading
  & All & Classic & Cold & Hot & SDOs & Cent\\
\hline % inserts single horizontal line 
This work & 30 & 19 & 9  & 10 & 10$^{(\star)}$ &  1\\ % inserting body of the table
LP		  & 53 & 33	& 16 & 17 &	10 & 10$^{(\bullet)}$\\
2MS		  & 33 & 21	& 6	 & 15 &	5  &  7\\	
\hline
Total	&  116 & 73	& 31 & 42 & 25 & 18\\	 
\hline %inserts single line
\end{tabular}
\footnotesize{Number of objects for the different dynamical groups (except the 2 Neptune{'}s resonant objects):
\emph{All} (the whole sample), \emph{Classic} (classical TNOs), \emph{cold}
(classical TNOs with inclinations$<4.5\,^{\circ}$), \emph{hot} (classical
TNOs with inclinations$>4.5\,^{\circ}$), \emph{SDOs} (scattered disk
objects), and \emph{Cent} (Centaurs). \emph{LP} is the ESO Large Program
(\cite{Boe02}; \cite{Peix}), and \emph{2MS} the Meudon Multicolor Survey (\cite{Doress_b}). $^{(\star)}$ We include the peculiar distant object 2000CR$_{105}$. $^{(\bullet)}$ We include the uncertain Centaur 2002CB$_{249}$.}
\end{table}

\clearpage

\begin{table*} \scriptsize
\caption{Correlations} % title of Table
\label{correl} % is used to refer this table in the text
\centering % used for centering table
\begin{tabular}{c c c c c c c c c c c} % centered columns (4 columns)
\hline\hline % inserts double horizontal lines
 & & & OWN & & & OWN+LP & & & OWN+LP+2MS &\\ % table heading
 & Correlated magnitudes& $\rho$ & SL(\%) & n & $\rho$ & SL(\%) & n & $\rho$ & SL(\%) & n\\
\hline % inserts single horizontal line 
Classicals  & \textbf{B-V vs. R-I} & 0.05 & 15.66 & 12 & \textbf{0.40} & \textbf{99.03} & \textbf{42} & 0.34 & 99.09 & 61\\
		 & \textbf{R-I vs. B-R} & 0.03 & 8.33 & 12 & \textbf{0.50 }& \textbf{99.87} & \textbf{42} & \textbf{0.38} & \textbf{99.70} & \textbf{61}\\
		 & \textbf{B-V vs. q }  & \textbf{0.59} & \textbf{95.13} & \textbf{12} & \textbf{0.37} & \textbf{98.24} & \textbf{43} & \textbf{0.40} & \textbf{99.83} & \textbf{64}\\

		 & \textbf{B-R vs. q}   & \textbf{0.55} & \textbf{92.96} & \textbf{12} & \textbf{0.46} & \textbf{99.71} & \textbf{43} & \textbf{0.46} & \textbf{99.98} & \textbf{64}\\
		 & \textbf{B-V vs. i}   &\textbf{-0.63} & \textbf{96.24} & \textbf{12} &\textbf{-0.48} & \textbf{99.83} & \textbf{43} &\textbf{-0.49} & \textbf{99.99} & \textbf{64}\\

		 & \textbf{B-R vs. i}   &\textbf{-0.57} & \textbf{94.32} & \textbf{12} &\textbf{-0.54} & \textbf{99.96} & \textbf{43} &\textbf{-0.54} & \textbf{$>$99.99} & \textbf{64}\\

		 & \emph{B-R vs. e}   &\emph{-0.38} & \emph{79.79} & \emph{12} &\emph{-0.31} & \emph{95.28} & \emph{43} &\emph{-0.29} & \emph{98.06} & \emph{64}\\

\hline
Hot		 & \textbf{B-V vs. V-R} & \textbf{0.75} &	\textbf{93.38} &\textbf{7} & \textbf{0.78} &	\textbf{99.97} &\textbf{23} & \textbf{0.56} &\textbf{99.94} &\textbf{38}\\
         & \textbf{B-V vs. R-I} & -0.02 &	3.52 &	 7 & \textbf{0.49} &\textbf{97.83} &	\textbf{23} & \textbf{0.41} &\textbf{98.38} &\textbf{36}\\
		 & \emph{V-R vs. R-I} &-0.32 &	67.53 &	10 & \emph{0.39} &	\emph{95.27} &\emph{27} & \emph{0.29} &\emph{92.79} &	\emph{40}\\
		 & \textbf{R-I vs. B-R} & -0.18 & 34.11 & 7 & \textbf{0.53} &\textbf{98.76} &\textbf{23} & \textbf{0.41} &	\textbf{98.50} &\textbf{36}\\
		 & \textbf{B-V vs. q}   & 0.29 &	51.60 &	 7 & \textbf{0.41} &\textbf{94.71} &	\textbf{23} & \textbf{0.38} &\textbf{97.96} &\textbf{38}\\
		 & \textbf{V-R vs. q}   & 0.18 & 40.20 &	10 & 0.29 &	86.08 &	27 & \textbf{0.46} &	\textbf{99.67} &	\textbf{42}\\

		 & \textbf{B-R vs. q}& 0.32 & 56.89 & 7 & 0.36 & 91.02 & 23 &\textbf{0.49}&\textbf{99.68}&	\textbf{38}\\
		 & \textbf{B-V vs. i}  &\textbf{-0.90} &\textbf{97.27} &\textbf{7} &\textbf{-0.72} &	\textbf{99.93} &\textbf{23} &\textbf{-0.47} &\textbf{99.57} &\textbf{38}\\
		 & \textbf{V-R vs. i}  &-0.51 &	87.45 &	10 &\textbf{-0.58} &\textbf{99.70} & \textbf{27} &\textbf{-0.40} &	\textbf{98.98} &\textbf{42}\\
		 & \textbf{B-R vs. i}   &\textbf{-0.82} &\textbf{95.77} &\textbf{7} &\textbf{-0.64} &	\textbf{99.72} &\textbf{23} &\textbf{-0.48} &\textbf{99.67} &\textbf{38}\\
		 & \emph{B-V vs. e}   &\emph{-0.50} &\emph{77.93} &	\emph{7}&\emph{-0.34} & \emph{89.33} &	\emph{23} &\emph{-0.25} &\emph{86.82} &	\emph{38}\\

\hline
Cold     & \textbf{B-V vs. V-R} &-0.60 &76.99 &	 5 &\textbf{-0.64} &\textbf{99.50}&\textbf{20}&-0.39 & 94.91 &	26\\
		 & \emph{V-R vs. R-I} &-0.33 &	65.42 &	 9 &\emph{-0.30} &\emph{83.42} &\emph{23} &\emph{-0.38} &	\emph{95.41} &\emph{29}\\
		 
\hline
SDOs	 & \textbf{B-V vs. V-R} & 0.43 &66.21 &	6 & \textbf{0.63} &\textbf{97.14} &	\textbf{13} & \textbf{0.76} &\textbf{99.87} &\textbf{19}\\
		 & \textbf{R-I vs. B-R} & 0.37 &	59.38 &	 6 & 0.30 &	70.11 &	13 & \textbf{0.62} &	\textbf{99.19} &\textbf{19}\\
		 & \textbf{B-V vs. q}  &-0.54 &77.52 &6 & \textbf{0.70} &\textbf{98.44} &\textbf{13} & 0.36 &	87.66 &	19\\
		 & \emph{R-I vs. q}& 0.17 &	36.26 & 9 & \emph{0.35}&\emph{86.46}&\emph{19}&0.21 &	69.30 &	25\\
		 & \textbf{B-R vs. q}&-0.60&82.03&6&\textbf{0.62}& \textbf{96.85}&\textbf{13} & 0.34 &	85.54 &	19\\
		 & \emph{B-V vs. i} & 0.49 &72.26 & 6 &\emph{0.49}&\emph{90.97}&\emph{13}&0.29 &	77.80 &	19\\
		 & \textbf{B-V vs. e} & 0.03 &5.09 & 6 &\textbf{-0.59}&\textbf{95.83}&\textbf{13} &-0.32 &81.97 &	19\\
		 & \textbf{R-I vs. e}&\textbf{-0.62}&\textbf{91.89}&\textbf{9} &\textbf{-0.46} &\textbf{95.11}&\textbf{19} &-0.31 &86.93 &	25\\
		 & \textbf{B-V vs. $H_{R}$} &-0.26 &43.47&	6 &\textbf{-0.77}&\textbf{99.27}& \textbf{13} &\textbf{-0.54} &\textbf{97.73}&\textbf{19}\\
		 & \textbf{B-R vs. $H_{R}$}&-0.31 &	51.78 &6 &\textbf{-0.81}&\textbf{99.49}& \textbf{13} &\textbf{-0.57} &\textbf{98.41} &\textbf{19}\\
\hline % inserts single horizontal line 
Cent	 & \textbf{B-V vs. V-R}&&&&0.55&88.02&9 &\textbf{0.67}&\textbf{99.07}&\textbf{16}\\
		 & \textbf{B-V vs. R-I}&&&&0.58&90.10&9 &\textbf{0.71}&\textbf{99.43}&\textbf{16}\\
		 & \textbf{V-R vs. Q}&&&&\textbf{0.65}&\textbf{96.15}&\textbf{11}&\textbf{0.58}& \textbf{98.37} & \textbf{18}\\
		 & \textbf{B-V vs. i}&&&&\textbf{-0.62}&\textbf{91.89}&\textbf{9}&\textbf{-0.53}&	\textbf{95.97}&\textbf{16}\\
		 & \textbf{B-R vs. i}&&&&\textbf{-0.68}&\textbf{94.67}&\textbf{9}&\textbf{-0.52}&	\textbf{95.39}&\textbf{16}\\
		 & \textbf{B-V vs. e}&&&&\textbf{0.72}&\textbf{95.73}&\textbf{9}&\textbf{0.74}& \textbf{99.59}&	\textbf{16}\\
		 
		 & \textbf{B-R vs. e}&&&&\textbf{0.58}&\textbf{90.10}&\textbf{9}&\textbf{0.67}&	\textbf{99.08} &\textbf{16}\\
		 & \textbf{V-R vs. a}&&&&\textbf{0.61}&\textbf{94.59}&\textbf{11}&\textbf{0.53}&	\textbf{97.04} &\textbf{18}\\
		 
		 &\textbf{B-V vs. $H_{R}$}&&&&\textbf{-0.72}&\textbf{95.73}&\textbf{9}&-0.18&50.57& 16\\
		      
		 & \textbf{B-R vs. $H_{R}$}&&&&\textbf{-0.73}&\textbf{96.19}& \textbf{9} &-0.27 &	70.83 &	16\\		  
\hline
All	  	 & \textbf{B-V vs. V-R}& 0.29 &	78.06 &	19 &\textbf{0.50}&\textbf{99.99}& \textbf{65} & \textbf{0.55} &\textbf{$>$99.99}&\textbf{99}\\
		 & \textbf{B-V vs. R-I}& 0.07 &	24.84 &	19 & \textbf{0.46}&\textbf{99.98}&\textbf{64} & \textbf{0.49}&\textbf{$>$99.99}&\textbf{96}\\

		 &\textbf{R-I vs. B-R}& 0.23 &66.51 &19&\textbf{0.56}&\textbf{$>$99.99}&\textbf{64}& \textbf{0.54} &\textbf{$>$99.99}& \textbf{96}\\
\hline
SDOs+hot & \textbf{B-V vs. V-R}& \textbf{0.55}&\textbf{94.30}&\textbf{13}&\textbf{0.70} &\textbf{$>$99.99}& \textbf{36}& \textbf{0.62} &\textbf{$>$99.99}& \textbf{57}\\
		 & \textbf{B-V vs. R-I} & $<$0.01 & 2.28 & 13 & \textbf{0.44}&\textbf{99.01}&\textbf{36} & \textbf{0.47}&\textbf{99.95}&\textbf{55}\\

		 & \textbf{R-I vs. B-R} & 0.17 &44.54 &	13 & \textbf{0.51}&\textbf{99.76}&\textbf{36} & \textbf{0.52} &\textbf{99.99} &\textbf{55}\\

		 & \textbf{V-R vs. i}&\textbf{-0.55}&\textbf{97.72}&\textbf{18}&\textbf{-0.44}&	\textbf{99.60}&	\textbf{44}&-0.35 &	99.44 &	 65\\
\hline
SDOs+Cent& \textbf{B-V vs. V-R}&0.57&83.84&7& \textbf{0.66}&\textbf{99.73}&\textbf{22}& \textbf{0.73}&\textbf{$>$99.99}& \textbf{35}\\
		 &\textbf{B-V vs. R-I}&0.04&6.97&7&0.47&96.88&22&\textbf{0.64}&\textbf{99.98}& \textbf{35}\\
	 	 & \textbf{V-R vs. R-I}&0.13&29.39&9&\emph{0.34}&\emph{92.27}&\emph{28}& \textbf{0.46} &	\textbf{99.64} &\textbf{41}\\
		 & \textbf{R-I vs. B-R}&0.32&56.89&7&\textbf{0.48}&\textbf{97.18}&\textbf{22}& \textbf{0.63}&	\textbf{99.98} &\textbf{35}\\
\hline %inserts single line
\end{tabular}

\footnotesize{Spearman correlation results of colors versus colors and
orbital parameters. OWN means our own data; OWN+LP our own data merged with the ESO Large Program data (\cite{Boe02}; \cite{Peix}); OWN+LP+2MS the OWN+LP data merged with the Meudon Multicolor Survey data
(\cite{Doress_b}), $\rho$= Spearman rank correlation, $SL$= Significance
level, $n$= Number of objects. Strongest correlations are in bold;
interesting or new, but weaker correlations are in italics.}

\end{table*}

\begin{table*} \scriptsize
\caption{Correlations and dependence on ``sizes".} % title of Table
\label{correlsizes1} % is used to refer this table in the text
\centering % used for centering table
\begin{tabular}{c c c c c c c c c c c} % centered columns (4 columns)
\hline\hline % inserts double horizontal lines
 & & & OWN & & & OWN+LP & & & OWN+LP+2MS &\\ % table heading
 & Correlated magnitudes& $\rho$ & SL(\%) & n & $\rho$ & SL(\%) & n & $\rho$ & SL(\%) & n\\
\hline % inserts single horizontal line 
Classicals & B-V vs. q  & 0.59 & 95.13 & 12 & 0.37 & 98.24 & 43 & 0.40 & 99.83 & 64\\
  &B-V vs. q($H_{R}>$6.2)&     &       & 2  & 0.28 & 84.33 & 27 & 0.33 & 93.04 & 32\\
  &B-V vs. q($H_{R}<$6.2)& 0.64& 94.37 & 10 & 0.52 & 95.50 & 16 & 0.49 & 99.40 & 32\\
  \hline
         & R-I vs. q   & 0.17 & 52.53 & 19 & 0.32 & 97.63 & 50 & 0.30 & 98.71 & 69\\
   &R-I vs. q($H_{R}>$6.2)&  0.71& 91.98 &  7 & 0.57 & 99.85 & 31 & 0.51 & 99.68 & 35\\
   &R-I vs. q($H_{R}<$6.2)& -0.10& 26.38 & 12 &-0.09 & 30.42 & 19 & 0.06 & 26.99 & 34\\
   \hline
         & B-R vs. q   & 0.55 & 92.96 & 12 & 0.46 & 99.71 & 43 & 0.46 & 99.98 & 64\\
   &B-R vs. q($H_{R}>$6.2)&      &       &  2 & 0.48 & 98.58 & 27 & 0.47 & 99.06 & 32\\
   &B-R vs. q($H_{R}<$6.2)&  0.55& 90.20 & 10 & 0.43 & 90.37 & 16 & 0.49 & 99.39 & 32\\

   \hline
         & R-I vs. e   & 0.17 & 54.82  & 19 &-0.21 & 86.62 & 50 &-0.17 & 84.83 & 69\\
   &R-I vs. e($H_{R}>$6.2)&  0.14& 27.36 &  7 & -0.41& 97.68 & 31 & -0.38& 97.28 & 35\\
   &R-I vs. e($H_{R}<$6.2)&  0.15& 38.26 & 12 &  0.13& 42.35 & 19 &  0.04& 19.76 & 34\\
   \hline

         & V-R vs. a   & -0.14 & 45.84 & 19 & 0.18 & 80.79 & 52 & 0.12 & 69.48 & 73\\
   &V-R vs. a($H_{R}>$6.2)& -0.43& 70.62 &  7 & -0.06& 26.64 & 33 & -0.02& 10.12 & 38\\
   &V-R vs. a($H_{R}<$6.2)&  0.03&  7.39 & 12 &  0.48& 96.01 & 19 &  0.21& 76.84 & 35\\
   \hline
         & B-R vs. a   & 0.08 & 21.92 & 12 & 0.09 & 43.80 & 43 & 0.14 & 73.85 & 64\\
   &B-R vs. a($H_{R}>$6.2)&		&		&  2 & -0.19& 67.17 & 27 & -0.11& 44.33 & 32\\
   &B-R vs. a($H_{R}<$6.2)&  0.25& 54.40 & 10 &  0.56& 97.04 & 16 &  0.29& 89.64 & 32\\
  \hline
        &V-R vs. $\psi$  &  0.20& 61.21 & 19 &  0.05& 27.68 & 52 &  0.05& 30.00 & 73\\
 &V-R vs. $\psi$($H_{R}>$6.2) &  0.14& 27.36 &  7 &  0.28& 88.87 & 33 &  0.35& 96.69 & 38\\
 &V-R vs. $\psi$($H_{R}<$6.2) &  0.10& 25.46 & 12 & -0.34& 84.70 & 19 & -0.28& 89.76 & 35\\
 \hline

        &B-R vs. $\psi$  &  0.20& 48.39 & 12 &$<$0.01& 1.44 & 43 & -0.02& 14.42 & 64\\
  &B-R vs. $\psi$($H_{R}>$6.2)&		&		&  2 &  0.06& 24.44 & 27 &  0.19& 70.87 & 32\\
  &B-R vs. $\psi$($H_{R}<$6.2)& -0.07& 15.85 & 10 & -0.23& 62.57 & 16 & -0.32& 92.06 & 32\\   	 
\hline % inserts single horizontal line 
Hot    &B-V vs. q    &  0.29& 51.60 & 7 & 0.41 & 94.71 & 23 & 0.38 & 97.96 & 38\\  
	 &B-V vs. q($H_{R}>$6.2)&    &       & 1 & 0.20 & 51.87 & 13 & 0.31 & 78.33 & 17\\
	 &B-V vs. q($H_{R}<$6.2)&0.77& 91.55 & 6 & 0.70 & 96.35 & 10 & 0.39 & 91.46 & 21\\
	 \hline	                                                                  
       &R-I vs. q  &   0.20 & 45.27  & 10& 0.23 & 75.89 & 27 & 0.19 & 76.46 & 40\\
   &R-I vs. q($H_{R}>$6.2) & 0.60 & 70.13 & 4& 0.37 & 84.55 & 16 & 0.20 & 61.21 & 19\\
   &R-I vs. q($H_{R}<$6.2) & 0.09 & 15.42 & 6& $<$0.01 & 2.30 & 11 & 0.13 & 45.04 & 21\\	 

	  \hline
	  &V-R vs. a        & -0.31 & 64.62 & 10 & 0.19 & 65.45 & 27 & 0.21 & 82.91 & 42\\
   &V-R vs. a($H_{R}>$6.2)   &      &       &  4 & -0.23& 61.99 & 16 & -0.16 & 51.09 & 20\\
   &V-R vs. a($H_{R}<$6.2)   & 0.37 & 59.38 &  6 & 0.67 & 96.66 & 11 & 0.49 & 97.41 & 22\\
   \hline
        &B-R vs. a   & $<$0.01 & $<$0.01 &  7 & 0.12 & 41.24 & 23 & 0.22 & 82.46 & 38\\
   &B-R vs. a($H_{R}>$6.2)&         &       &  1 & -0.13& 33.84 & 13 & -0.09& 26.85 & 17\\		
   &B-R vs. a($H_{R}<$6.2)&    0.60 & 82.03 &  6 & 0.79 & 98.28 & 10 & 0.59 & 99.15 & 21\\
   \hline
  Cold      &R-I vs. q  &    0.15 & 32.86 &  9 & 0.25 & 75.53 & 23 & 0.27 & 84.35 & 29\\
   &R-I vs. q($H_{R}>$6.2)&         &       &  3 & 0.56 & 96.29 & 15 & 0.60 & 98.06 & 16\\
   &R-I vs. q($H_{R}<$6.2)&   -0.14 & 25.06 &  6 & -0.45& 76.86 &  8 &-0.30 & 69.60 & 13\\

   \hline
           &V-R vs. a&   -0.17 & 36.26 &  9 & 0.35 & 91.29 & 25 & 0.09 & 38.31 & 31\\
   &V-R vs. a($H_{R}>$6.2)&   -0.50 & 52.05 &  3 & 0.41 & 90.05 & 17 & 0.42 & 91.52 & 18\\
   &V-R vs. a($H_{R}<$6.2)&   -0.20 & 34.53 &  6 & 0.17 & 34.08 &  8 & -0.30& 69.60 & 13\\
   \hline

\end{tabular}

\footnotesize{Correlations results of colors versus orbital parameters and
versus $\psi$, a measure of the ``average Collisional Energy'' (\cite{Opik76})
for classical objects. We correlate the whole group and two different
subgroups of different ``sizes'' with $H_{R}>$6.2, and $H_{R}<$6.2. OWN means
our own data, OWN+LP means our own data merged with the ESO Large Program
data (\cite{Boe02}; \cite{Peix}), OWN+LP+2MS means the OWN+LP data merged with the Meudon
Multicolor Survey data (\cite{Doress_b}), $\rho$= Spearman rank correlation,
$SL$= Significance level, $n$= Number of objects.}

\end{table*}

\begin{table*}
\caption{Remarkable or interesting correlations of computed spectral gradients.} % title of Table
\label{Grt_correl} % is used to refer this table in the text
\centering % used for centering table
\begin{tabular}{c c c c c c c c c c c} % centered columns (4 columns)
\hline\hline % inserts double horizontal lines
 & & & OWN & & & OWN+LP & & & OWN+LP+2MS &\\ % table heading
 & Correlated magnitudes & $\rho$ & SL(\%) & n & $\rho$ & SL(\%) & n & $\rho$ & SL(\%) & n\\
\hline % inserts single horizontal line 
Classicals & Grt vs. i	&	-0.28	&	77.11	&	19	&	-0.49	&	99.95	&	52	&	-0.47	&	99.99	&	73	\\
& Grt vs. i (H$_{R}>6.2$)	&	-0.68	&	90.35	&	7	&	-0.61	&	99.94	&	33	&	-0.66	&	99.99	&	38	\\
& Grt vs. i (H$_{R}<6.2$)	&	-0.17	&	42.29	&	12	&	-0.37	&	88.47	&	19	&	-0.36	&	96.48	&	35	\\
& Grt vs. $\psi$&	0.12	&	37.68	&	19	&	-0.03	&	16.76	&	52	&	-0.08	&	48.26	&	73	\\
& Grt vs. $\psi$ (H$_{R}>6.2$)	&	-0.14	&	27.36	&	7	&	0.13	&	55.39	&	33	&	0.16	&	66.45	&	38	\\
& Grt vs. $\psi$ (H$_{R}<6.2$)	&	-0.20	&	49.88	&	12	&	-0.41	&	91.71	&	19	&	-0.40	&	97.90	&	35	\\

\hline
Hot & Grt vs. Q	&	-0.24	&	52.17	&	10	&	-0.04	&	17.01	&	27	&	-0.03	&	16.00	&	42	\\
& Grt vs. Q (H$_{R}>6.2$)	&	-0.40	&	51.16	&	4	&	-0.30	&	75.93	&	16	&	-0.32	&	83.34	&	20	\\
& Grt vs. Q (H$_{R}<6.2$)	&	0.26	&	43.47	&	6	&	0.54	&	91.01	&	11	&	0.40	&	93.34	&	22	\\
& Grt vs. a	&	-0.08	&	18.68	&	10	&	0.11	&	40.76	&	27	&	0.21	&	81.33	&	42	\\
& Grt vs. a (H$_{R}>6.2$)	&	-0.40	&	51.16	&	4	&	-0.16	&	46.90	&	16	&	-0.09	&	30.59	&	20	\\
& Grt vs. a (H$_{R}<6.2$)	&	0.66	&	85.83	&	6	&	0.61	&	94.59	&	11	&	0.57	&	99.12	&	22	\\
 
\hline
SDOs & Grt vs. e	&	-0.32	&	62.96	&	9	&	-0.49	&	96.28	&	19	&	-0.27	&	81.78	&	25	\\
& Grt vs. H$_{R}$	&	-0.52	&	85.61	&	9	&	-0.64	&	99.34	&	19	&	-0.34	&	90.86	&	25	\\
\hline
Cent & Grt vs. i	&		&		&		&	-0.31	&	67.16	&	11	&	-0.31	&	79.67	&	18	\\
& Grt vs. e	&		&		&		&	0.39	&	78.36	&	11	&	0.56	&	98.01	&	18	\\
& Grt vs. H$_{R}$	&		&		&		&	-0.51	&	89.26	&	11	&	-0.26	&	72.21	&	18	\\
\hline
All & Grt vs. q	&	0.13	&	52.26	&	30	&	0.32	&	99.66	&	83	&	0.34	&	99.97	&	116	\\
& Grt vs. i	&	-0.47	&	98.78	&	30	&	-0.48	&	100.00	&	83	&	-0.42	&	100.00	&	116	\\
& Grt vs. e	&	-0.28	&	86.45	&	30	&	-0.28	&	98.85	&	83	&	-0.27	&	99.68	&	116	\\
& Grt vs. H$_{R}$	&	-0.63	&	99.93	&	30	&	-0.28	&	98.91	&	83	&	-0.17	&	93.72	&	116	\\
\hline
 SDOs+hot & Grt vs. q	&	-0.03	&	9.48	&	19	&	0.31	&	96.06	&	46	&	0.38	&	99.78	&	66	\\
& Grt vs. i	&	-0.41	&	91.73	&	19	&	-0.37	&	98.75	&	46	&	-0.27	&	97.13	&	66	\\
& Grt vs. H$_{R}$	&	-0.57	&	98.38	&	19	&	-0.40	&	99.21	&	46	&	-0.26	&	96.39	&	66	\\
\hline
SDOs+Cent & Grt vs. H$_{R}$	&	-0.65	&	94.83	&	10	&	-0.48	&	99.02	&	30	&	-0.19	&	76.87	&	42	\\
\hline
\end{tabular}

\footnotesize{Interesting correlations of the spectral gradient (Grt) versus
$H_{R}$ and orbital parameters discussed in text. The spectral gradient,
Grt(\%$/100 nm$), was calculated by fitting over the available color
intervals per object. OWN means our own data, OWN+LP means our own data
merged with the ESO Large Program data (\cite{Boe02}; \cite{Peix}), OWN+LP+2MS means the
OWN+LP data merged with the Meudon Multicolor Survey data (\cite{Doress_b}),
$\rho$= Spearman rank correlation, $SL$= Significance level, $n$= Number of
objects.}

\end{table*}

\end{document}